# Towards electron encapsulation. Polynitrile approach.

Ilya A. Shkrob [a)] and Myran C. Sauer, Jr.

*Chemistry Division, Argonne National Laboratory, 9700 S. Cass Ave, Argonne, IL 60439*

## Abstract

This study seeks an answer to the following question: Is it possible to design a supramolecular cage that would "solvate" the excess electron in the same fashion in which several solvent molecules do that co-operatively in polar liquids? Such an "encapsulated electron" would be instrumental for structural and dynamics studies of electron solvation and might be useful for molecular electronics. Two general strategies are outlined for "electron encapsulation," viz. electron localization using polar groups arranged on the (i) inside of the cage or (ii) outside of the cage. The second approach is more convenient from the synthetic standpoint, but it is limited to polynitriles. We demonstrate, experimentally and theoretically, that this second approach faces a formidable problem: the electron attaches to the nitrile groups forming molecular anions with bent *C-C-N* fragments. Since the energy cost of this bending is very high, for dinitrile anions in *n*-hexane, thebinding energies for the electron are very low and for mononitriles, these binding energies are lower still, and the entropy of electron attachment is anomalously small. Density functional theory modeling of electron trapping by mononitriles in *n*-hexane suggests that the mononitrile molecules substitute for the solvent molecules at the electron cavity, "solvating" the electron by their methyl groups. We argue that such "solvated electrons" resemble multimer radical anions in which the electron density is shared (mainly) between *C 2p* orbitals in the solute/solvent molecules, instead of existing as cavity electrons. The way in which the excess electron density is shared by such molecules is similar to the way in which this sharing occurs in large di- and poly- nitrile anions, such as 1,2,4,5,7,8,10,11-octacyano-cyclododecane$^-$. The work thus reveals limitations of the concept of "solvated electron" for organic liquids: it might be impossible to draw a clear line between such species and a certain class of radical anions. It also demostrates the feasibility of "electron encapsulation."

______________________________________________________________

[a)] Author to whom correspondence should be addressed; electronic mail: shkrob@anl.gov.

## 1. Introduction.

In liquids that consist of molecules with no electron affinity, the excess electron sometimes exists as a metastable "solvated electron" in which the electron wavefunction occupies an interstitial cavity between several solvent molecules. [1,2] In polar liquids, such as water and alcohols, this cavity is lined by $-O^{\delta-}H^{\delta+}$ groups pointing towards the center of the cavity (Figure 1a). [3] Stabilization of the cavity electron through Coulomb attraction to permanent dipoles in the solvent molecules is opposed by repulsion due to the Pauli exclusion. To a first approximation, such a species can be regarded as a single quantum particle (the excess electron) in a classical potential well; this electron is treated separately from the valence electrons in the solvent molecules. The models of "solvated electron" differ chiefly in how this classical potential is implemented and various approximations made in the treatment of electron dynamics. The current state-of-the-art quantum mechanics - molecular dynamics (QM/MD) models of electron solvation [3] treat the liquid (usually, water) as an assembly of flexible, polarizable molecules (that are treated classically) and rigorously treat nonadiabatic dynamics of the electron in the fluctuating classical potential, which is prescribed (the electron is treated fully quantum mechanically).

Is the key simplification made in such one-electron models (the separability of a single quantum particle from the rest of the system that is treated classically) always sufficient to capture the observed properties of the "solvated electron?" The probable answer is no. A specific concern is that there might be a significant transfer of the excess electron density into the frontier orbitals of groups at the cavity wall. If such a transfer is not negligible, the core of the "solvated electron" should be more properly viewed as a multimer radical anion rather than a cavity electron. In our previous study, [4] we provided several arguments that such is indeed the case for the excess electron in liquid ammonia ("ammoniated electron") where the spreading of the negative charge onto *N 2p* orbitals of ammonia molecules in the first and second solvation shells is considerable, as suggested by density functional theory (DFT), [4] *ab initio* [5] calculations, and $^{14}$N and $^{1}$H NMR spectroscopy. [6] In practice, distinguishing between the cavity electron (in which most of the excess electron density is contained in a void between the solvent molecules) and a multimer radical anion (in which most of the excess electron density is in the frontier orbitals of several solvent molecules) may not always be possible, because there is no sharp boundary between these two modes of electron solvation. [7]

The dilemma of "cavity electron" vs. "multimer radical anion" is especially vexing for organic liquids of no or low polarity, because permanent dipoles in the molecules are either lacking or weak, and the attractive potential originates entirely or mainly through the bond polarization. In water, the potential well for the "solvated electron" is > 1.5 eV deep [2,3,4] and the electron wavefunction is largely contained within the potential well, with little electron density spilled over the outside (which justifies the use of the "solvated electron" picture for this liquid). By contrast, in liquid alkanes (such as *n*-hexane), the binding energy of the "solvated electron" is only 180-200 meV, [8,9] and the localization radius *a* of the exponentially decaying electron density $4\pi r^2 \rho(r) \propto \exp(-2r/a)$ is ca. 4-5 Å. [9] This localization radius is larger than the radius of the cavity (3.2-3.6 Å), as determined from magnetic resonance data [10] for trapped electrons in vitreous alkanes and the reaction volume for electron localization in liquid and supercritical alkanes. [8] With the electron wavefunction that diffuse, neglecting the interaction between the excess electron and valence electrons in the solvent molecules is poorly justified. Indeed, the existing models for solvated electrons in liquid alkanes have

4.

difficulty explaining the origin of the trapping potential.[11,12,13] The expected degree of polarization of alkane molecules appears to be insufficient to bind the excess electron. This persistent problem suggests that one-electron models may not suffice for the alkanes: the origin of the trapping potential might be the increased electron density in the frontier *C 2p* orbitals that weakens the *C-C* and *C-H* bonds and makes the alkane molecule more polarizable. The closer one examines such a "solvated electron," the more it resembles a radical anion.

**1.1. Electron encapsulation.**

The same question can be brought to a sharp focus by the following *Gedanke* experiment ("electron encapsulation"). Suppose that the entire first solvation shell of the solvated electron is replaced by a single supramolecular structure (the "cage") that has the internal cavity lined by polar groups (the conceptual drawing of such a cage is given in Figure 1b). The cage is suspended in a liquid with very low binding energy for the excess electron (such as an alkane or rare gas liquid). Assuming that the cage traps the excess electron, what is the result of this "electron encapsulation?" Should one regard the resulting species as a kind of "solvated electron" or as a large molecular anion? In which way and by what criteria is it possible to distinguish between these two extremes?

The "electron encapsulation" is more than a thought experiment; it is an experimental possibility. Electron trapping by hydrogen-bonded clusters of 2-4 alcohol molecules in nonpolar liquids has been studied by several groups,[14] including these authors.[15] We have recently demonstrated[15,16] that, due to the large dipole moment of acetonitrile (3.9-4 D vs. 1.6 D for a typical alcohol), a single acetonitrile molecule can trap the electron in *n*-hexane and *iso*-octane, apparently without the formation of a molecular anion. In both of these cases, one obtains a reducing species with absorption spectra and chemical properties that are indistinguishable from those of "solvated electrons" in neat liquids. Electron trapping by interstitial cavities that have well-defined, fixed geometry occurs in single crystals of carbohydrates (such as sucrose) [17] and in electrides. [18] In the former, trapped electrons with regular ligand geometry can be observed using electron paramagnetic resonance (EPR) and optical spectroscopy. These observations suggest that "electron encapsulation" by a rigid supramolecular structure is possible.

What macrocycles could serve as such an electron-trapping cage? The molecule should satisfy certain conditions that would preclude the formation of a molecular anion with the negative charge localized on the scaffolding. In particular, it should contain no

- aromatic rings and *C=C* double bonds;
- *C=O* groups (with the possible exception of amide groups), halogen atoms (with the possible exception of F), and other electrophilic groups (such as nitro groups);
- protons (e.g., quaternized amino groups) and other groups that rapidly react with the electron by, e.g., proton transfer.

In order to deliver the electron to such a cage, one should be able to disperse these cages in solvents for which electron binding is very weak, such as alkanes or hexamethyl phosphoramide. [19] The latter has the advantage of being a strongly polar liquid. The electron affinity of the cage should be larger than that of the solvent itself, as otherwise the electron would stay in the solvent.

The most difficult requirement to meet is that the polar groups should be arranged in such a way that the dipoles point towards the cavity center, with the positive charge closer to the center of the cavity (Figure 1b). In the vast majority of synthetic macrocycles (such as crown ethers, azacrowns, cryptands, etc.) the opposite is the case,

6.

since these macrocycles are *cation* rather than *anion* complexants. Thus, the search for a feasible cage for "electron encapsulation" naturally leads one to *anion* (especially, halide) receptors, [20,21] because the halide anions are solvated in a fashion that resembles electron solvation. The field of synthetic anion receptors is relatively new, and the choice of possible structures that fulfil the demanding criteria given above is limited. Most of the synthetic anion receptors are polyammonium macro(bi)cycles that stabilize the interstitial halide ($X^-$) anion through Coulomb attraction to quaternized amino groups placed on ring, tetrahedral, or octahedral patterns (see Figures 1S, (a) to (d) in the Supplement). [20,21] In some of these macrobicycles (e.g., for the octaazacryptand shown Figure 1S(b)), there is additional stabilization through the formation of $X^-...H$-$N$ bonds. [22] While recent experiments of Wishart and Neta [23] on solvated electrons in ionic liquids suggest that electron trapping by 3-4 fully functionalized $NR_4^+$ cations might be possible, such "encapsulated electrons" would be the synthetic analogs of F-centers in ionic crystals rather than "solvated electrons" in common organic liquids. Other, more exotic, classes of anion receptors (such as, for example, amide macrocycles, [24] Lewis acids [e.g., silacrowns], [20] and porphyrin-inspired [25] rings) do not fulfil the requirements stipulated above. Our exploratory studies indicate that without quaternization, azacrown, katapinand, and azacryptand macro(bi)cycles do not trap electrons in alkane solvents. The small negative enthalpy of such trapping, which can only occur via electron solvation by >$N$-$H$ groups (that have small dipole moments < 0.5 D), is insufficient to cancel a large loss in the entropy.

This examination suggests that the electron-trapping cavity can be lined only by highly polar groups, such as *OH* groups, in order to provide a sufficiently deep well for the electron. The design and synthesis of such a cage would be challenging, because the *OH* groups have to be arranged *inside* the cavity, whereas the natural preference for such groups is at the outside. While there are structural motifs that can be explored (such as cyclodextrins and their analogs), in this paper we focus on a different approach: using outward pointing >*C-CN* groups as electron-binding dipoles, as shown in Figure 1c. Since the positive charge in this dipole is on the carbon, such dipoles, when placed outside, still point towards the center of the cavity. The electron "solvation" is carried out by >*CH$_2$* and/or >*NH* groups at the cavity walls. Since the dipole moment of the *-CN* group is large (3.9-4 D), such a placement still facilitates electron trapping by the cavity. Even in neat liquid acetonitrile [16,26] and $(CH_3CN)_n^-$ clusters, [27,28] the electron is solvated by *methyl* groups rather than cyano groups of the acetonitrile which point away from the cavity. A similar situation occurs for electrons in amides, where polar *C=O* groups also point away from the cavity. [19,29] Thus, one can readily envision macro(bi)cycles with counter pointing cyano groups (such as the structures shown in Figures 2S, (a) and (b) in the Supplement). The electron occupies the cavity formed by methylene and amino groups; the stabilization of this cavity electron occurs via electrostatic binding to several *CN* dipoles placed outside of the cavity (Figure 1c).

The experimental implementation of this scheme gave an unexpected result (section 3): placing just two cyano groups on the aliphatic scaffolding results in the formation of a molecular anion instead of the anticipated "solvated electron" like species. Yet such molecular anions are not dissimilar to "solvated electrons" that occur in neat alkanes and nitriles as suggested by our DFT calculations (section 4). Our result points to the limitations of the concept of "solvated electron" as applied to organic molecules. The program intended as synthetic re-creation of "cavity electron" brings instead the realization that such a species might be a theoretical abstraction. Before these results are examined, we briefly review what is known about electron trapping by nitriles and dinitriles.

### 1.2. Electron trapping in nitriles.

Electron localization in solid nitriles ($H(CH_2)_nNC$) and dinitriles ($NC(CH_2)_nNC$) yields three kinds of species. Dissociative attachment (which is the main reaction in vitreous butyronitrile) [16,30] yields a residual *C*-centred alkyl radical and $CN^-$ anion, whereas nondissociative attachment yields either a monomer or a dimer radical anion, depending on the crystalline structure and the temperature. [31] These anions can be observed by their transient absorption spectra [31,32] or distinctive features in their Electron Paramagnetic Resonance (EPR) spectra. [32a] Shkrob et al. [33] and others [26,31] found that acetonitrile molecules in the dimer and the monomer anions have similar geometry, with the *C-C-N* angle of 130°. In the centrosymmetrical dimer anion, there is a 3-electron-2-center bond between the two cyanide carbons with the *C-C* distance of 1.65 Å. [33] Other nitriles are expected to form anions with the same structural motif. Radiolysis of solid succinonitrile (*n*=2), either in cubic (plastic) or monoclinic phases at 77 K results in the formation of a dimer anion that absorbs at 540 nm; [32a] irradiation of a room temperature cubic crystal yields a monomer anion that absorbs at 450 nm and decays in 230 ns via the loss of $CN^-$. [32b] Such a dissociation can also be stimulated by absorption of visible light, and it was observed for all acetonitrile anions, both in solid [31,33] and in neat liquid. [16] Crystalline adiponitrile is also known to yield dimer radical anion in low-temperature radiolysis. [32a] Apart from these observations, there have been no reports of molecular anions of nitriles and dinitriles, in any media.

In liquid acetonitrile, the are two forms of localized electron that are in equilibrium with each other: the dimer anion, which has an absorption spectrum similar to that for the dimer anion in solid *α*-acetonitrile, [16,26,34] and "solvated electron" that absorbs in the infrared. [16,26] These two pathways of electron localization are also found in large acetonitrile clusters. [27] The electron is solvated by methyl groups of acetonitrile, as

9.

suggested by QM/MD simulations [28a] and density functional theory (DFT) [16] and *ab initio* calculations. [28b] In the liquid acetonitrile, the dimer anion is 450 meV more stable than the "solvated electron," but the formation of this anion requires bending of the *C-C-N* group, which results in a large barrier towards *C-C* bonding and dimerization. When acetonitrile molecules are isolated in dilute *n*-hexane solution, this dimerization cannot occur, yet a species is formed that has a spectrum resembling that of a "solvated electron" in neat *n*-hexane. This trapped-electron species has an electron binding energy that is only ca. 0.2 eV lower than the binding energy of the electron in a solvent trap, and a standard entropy of trapping that is extremely small, ca. -15 J/mol.K. [15,16] These observations led us to suggest that acetonitrile monomer does not form a molecular anion in *n*-hexane. [15] Rather, the methyl group of the acetonitrile molecule is at the wall of the electron cavity together with the methyl/methylene groups of the solvent molecules. The cavity electron is dipole-bound to the *CN* dipole (with *the C-C-N* angle of 180°), which results in a more stable trap. A simple electrostatic calculation given in ref. 16 suggests that such a localization mechanism would be energetically feasible. In other words, for acetonitrile in *n*-hexane the dilemma of "molecular anion" vs. "solvated electron" is resolved in favor of the latter, because the formation of a molecular anion requires energetically costly bending of the *C-C-N* fragment whereas the "solvation" requires minimal change in the geometry. This picture is further refined in sections 4.1 and 4.4. The important point is that it is not obvious *a priori* whether the dinitriles would follow the same pattern since the electron affinity for such molecules, while still negative, is less negative than that for acetonitrile. Only experiment can establish what is the mode of electron localization for such molecules. To this end, we used two-laser two-color time-resolved d. c. photoconductivity, as explained in the next section.

## 2. Experimental.

All chemicals were obtained in the purest form available from Aldrich, and used without further purification; *n*-hexane (99+%) was passed through a 2 m long column filled with activated silica gel prior to use. The structures of (di-)nitriles are given in Figure 3S in the Supplement. Other figures and sections with designator "S" can also be found therein. In the following, CH2N stands for *trans*-1,4-cyclohexane dinitrile (Figure 3S).

All measurements of electron dynamics were carried out using $N_2$-saturated solutions of *n*-hexane. The typical (impurity limited) lifetime of the electron was 300 ns. A representative set of the decay kinetics is given in Figure 2a (for glutaronitrile). The conductivity setup was the same as described in our previous publications. [9,15,16] The solution was placed in a 2 cm path optical cell with suprasil windows and ionized using two 248 nm photons from a KrF excimer laser (21 ns fwhm pulse, < 0.1 $J/cm^2$). The second laser (Nd:YAG ,1064 nm, 6 ns fwhm, < 1 $J/cm^2$) was used to photoexcite trapped electron (or molecular anion), thereby injecting a quasfree electron into the conduction band of the liquid at a certain delay time $t_L$ after the ionization event (50-500 ns; the time jitter between the two laser pulses was < 3 ns). This photoexcitation yields a short-lived "spike" in the conductivity, as shown in Figure 2b, where the 248 nm induced signal has been subtracted from the signal obtained with both lasers. The decay of this spike represents re-establishment of the electron equilibria. The two laser beams propagated in the opposite directions and the 1064 nm beam completely enveloped the collimated 248 nm beam inside the conductivity cell. An electric field of 6-8 kV/cm was applied to two flat platinum electrodes spaced by 6.5 mm. The cell was placed in a metal jacket; the temperature of the photolyzed solution was controlled by circulating water containing antifreeze through this jacket. The photoconductivity signal $\kappa(t)$ was amplified and

11.

recorded on a transient digitizer with time resolution < 2 ns. The conductivity is given in units of nS/cm ($10^{-7}$ $\Omega^{-1}$ $m^{-1}$); if not specified otherwise, the data were taken at 291 K (the "standard conditions" refer to 295 K). To determine the signal $\Delta\kappa(t)$ induced by the 1064 nm light, the Nd:YAG laser was pulsed on and off while the 248 nm laser was pulsed for every shot, and the difference signal was defined as $\Delta\kappa(t) = \kappa_{on}(t) - \kappa_{off}(t)$ (Figure 2b).

## 3. Results.

Typical decay kinetics for the photoconductivity signal observed after two 248 nm photon ionization of *n*-hexane solutions (the (di-)nitriles do not absorb at 248 nm) are shown in Figure 2a. The signal is from a mobile, quasifree electron $e_{qf}^-$, which is in thermodynamic equilibrium with the solvated electron

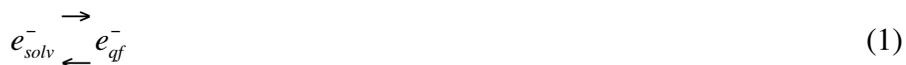

$$e_{solv}^- \underset{\leftarrow}{\overset{\rightarrow}{\phantom{xx}}} e_{qf}^- \qquad (1)$$

This equilibrium is strongly shifted towards the left side: the lifetime of the localized (solvated) electron before it is thermally emitted back into the conduction band of the solvent is 8-10 ps, [9] whereas the trapping time of the quasifree electron is only 20-30 fs, i.e., it occurs even faster than the relaxation of its momentum (at 295 K). [9,35] The standard enthalpy of reaction (1) is ca. 0.2 eV. [8,9] Since the mobility of the quasifree electron is 10-100 $cm^2$/Vs, whereas that of the solvated electron is only $10^{-4}$-$10^{-3}$ $cm^2$/Vs, the conductivity signal is due to the small equilibrium fraction of the quasifree electrons. Due to the rapid settling of equilibrium reaction (1), this fraction tracks the concentration of solvated electrons, $e_{solv}^-$. The apparent life-time of the conductivity signal is controlled by electron-accepting impurity, such as traces of oxygen and fluorocarbons. The conductivity signal decays exponentially as $\kappa(t) = \kappa_0 \exp(-k_0 t) + \kappa_i$, where $k_0$ is rate constant for pseudo first order (irreversble) electron attachment to electron-scavenging impurity and $\kappa_\iota$ is the conductivity signal from ions. The radical cation of *n*-hexane rapidly deprotonates;[36] the resulting cation has low mobility and makes negligible contribution to the overall conductivity signal. On the time scale of our kinetic observations (< 2 µs), the loss of the electrons/ions due to their recombination in the bulk and migration in the electric field and discharge at the electrodes is negligible (the typical electron concentration was just a few nM) .

In the presence of (di-)nitriles, the electron reversibly reacts with the solute (*S*) yielding a product $\{e^- : S\}_{solv}$ (no specific structure is attributed to this product at this point):

$$e^-_{solv} + S \rightleftarrows \{e^- : S\}_{solv} \qquad (2)$$

For 1,1-dinitriles (such as malono- and dimethylmalono- nitriles) the equilibrium is strongly shifted towards the product, so that backward reaction (2) is too slow to occur on the observation time scale. It is also possible that this attachment is dissociative, as occurs for electron attachment to the nitriles in *polar* solvents (the free energy gain due to the *C-C* bond dissociation is sufficiently great to exceed the large free energy loss due to the solvation of the small CN⁻ anion in the nonpolar medium).

Careful analysis of the decay kinetics suggests that for 1,2-, 1,3-, and 1,4- di-nitriles the product $\{e^- : S\}_{solv}$ reacts (with diffusion-controlled reaction rate) with another solute molecule yielding a *C-C* bridged dimer anion $S_2^-$:

$$\{e^- : S\}_{solv} + S \rightleftarrows S_2^- \qquad (3)$$

The structure of this dimer anion should be similar to those of acetonitrile, succinonitrile, and adiponitrile dimer anions as observed by absorption spectroscopy and EPR [31,32,33] and suggested by quantum chemistry calculations. [26,33] Reaction (3) shifts equilibrium (2) to the right side which makes it necessary to take it into account in order to determine the equilibrium parameters of reaction (2). For glutaro- and succino- nitrile (Figures 2 and 3, respectively), the rate constants for dimerization reaction (3) are close to those for diffusion-controlled reactions in $n$-hexane; for adiponitrile this reaction is ca. 10 times slower (Table 1). For mononitriles (see Figure 4 for butyronitrile), no evidence for the occurrence of reaction (3) has been found, as explained in refs. 15 and 16. For these solutes, equilibrium is reached within the duration of the 248 nm laser pulse, so the amplitude of the exponentially extrapolated prompt conductivity signal $\kappa_0$ decreases as $(1 + K_{eq}[S])^{-1}$, where $K_{eq}$ is the equilibrium constant of reaction (2); [15] the apparent rate of electron attachment to impurity also decreases, since the apparent mobility of the charge carrier decreases in the same proportion as the equilibrium fraction of solvated (and, therefore, quasifree) electrons. [9,15] Photoexcitation of $\{e^- : S\}_{solv}$ by infrared light causes electron detachment and injection into the conduction band of the solvent [15]

$$\{e^- : S\}_{solv} \xrightarrow{h\nu} S + e^-_{qf} \qquad (4)$$

The released quasifree electron then undergoes reactions (1) to (3). Since the dimer anions have low cross-section for photodetachment at 1064 nm, [31,33] the initial magnitude of the signal $\Delta\kappa_0 = \Delta\kappa(t = t_L)$ induced by the 1064 nm light is proportional to the instantaneous concentration $e^-_{solv}$ of solvated electrons in the reaction mixture immediately prior to the 1064 nm photon excitation at $t = t_L$. For mononitriles, the time profile of this prompt signal follows the $\kappa(t_L) - \kappa_i$ kinetics (of the electron decay) exactly. For dinitriles, the latter decay kinetics deviates at longer delay times (two

14.

examples of this behavior are shown in Figures 2b, for glutaronitrile, and Figure 4S, for CH2N) thereby indirectly supporting the occurrence of reaction (3). For mononitriles, settling of equilibrium (2) is very fast, and the conductivity signal $\Delta\kappa(t)$ induced by the 1064 nm light simply follows the time profile of the 1064 nm laser pulse. For dinitriles, the reactions are sufficiently slow that settling of the equilibrium following electron photodetachment (4) (or the initial photoionization) can be time resolved, and that allowed us to estimate the rate constants of forward and backward reactions (2) and reaction (3) directly. Relevant formulas are derived in the Appendix given in the Supplement.

In order to derive the thermodynamics parameters of reaction (2) the families of decay kinetics for different concentrations $[S]$ of the solute at several temperatures between –10 ºC and +50 ºC were analyzed (Figures 3, 4S, and 5S for succinonitrile, CH2N, and adiponitrile, respectively). The corresponding van't Hoff plots (Figure 5) of the equilibrium constant $K_{eq} = k_2/k_{-2}$ gave the standard enthalpy $\Delta H^0$ and entropy $\Delta S^0$ of reaction (2) at 295 K. For some solutes (especially, at elevated temperatures), the settling of the equilibrium was too rapid to resolve in time; in such instances the equilibrium fraction of $e^-_{solv}$ was estimated from the extrapolated $\kappa(t)$ kinetics, and the equilibrium constant $K_{eq}$ was determined from a linear correlation plot: $\kappa_0([S]=0)/\kappa_0 = 1 + K_{eq}[S]$. In some instances, we were still able to estimate the settling time of reaction (2) from the "tails" the 1064 nm induced kinetics, as explained in ref. 15 and the Appendix, because the 1064 nm pulse is much shorter than the 248 nm pulse (an example of such analysis for glutarontrile solution is shown in Figures 2b and 2c).

These kinetic and thermodynamic data can be summarized as follows: 1,1-dinitriles such as malononitrile and dimethylmalononitrile react with the electron rapidly

15.

and irreversibly (on our time scale) with rate constants of (0.63±0.03) and (1.04±0.05) x$10^{12}$ $M^{-1}$ $s^{-1}$, respectively (at room-temperature). For comparison, reaction constants for electron scavenging by nitrobenzene, $SF_6$, $CCl_4$, $CO_2$, and $N_2O$ (all of which are efficient electron scavengers) are 2.0, 2.0, 1.2, 1.8, and 1.5 x$10^{12}$ $M^{-1}$ $s^{-1}$, respectively. [37] For succinonitrile, forward reaction (2) is also very rapid (Table 1); the rate constant is ca. 2.2x$10^{12}$ $M^{-1}$ $s^{-1}$. These rapid reactions suggest highly exoergic electron attachment and indeed for succinonitrile the standard enthalpy of reaction (2) is large, ca. –0.68 eV, and the absolute free energy is the greatest among the dinitriles for which the thermodynamic data were obtained (Table 1). For other dinitriles, including malononitrile, forward reaction (2) is considerably slower, (4-7)x$10^{11}$ $M^{-1}$ $s^{-1}$. Such relatively slow electron-attachment reactions were previously observed for solutes with low electron affinity, such as styrene (4.2x$10^{11}$ $M^{-1}$ $s^{-1}$), $\alpha$–methylstyrene (6.3x$10^{11}$ $M^{-1}$ $s^{-1}$), and methylpentafluorobenzene (3.9x$10^{11}$ $M^{-1}$ $s^{-1}$),[37] all of which reversibly attach the electron and form molecular anions [38] (whose structure is well known from matrix isolation EPR and absorption spectroscopy). The standard free energies of electron attachment to styrene, $\alpha$–methylstyrene, and difluorobenzene are –0.56, -0.51, and –0.33 eV, [38,39] respectively, which are significantly more negative than the free energies given in Table 1. The latter are close to the free energies of electron attachment to butadiene (-0.2 eV) as estimated by Holroyd [38] from high-pressure conductivity data. The standard enthalpies of electron attachment to styrene and $\alpha$–methylstyrene (-1.1 eV) [39] are also much lower than those for the (di-)nitriles in Table 1. Electron scavenging by the mono- and di- nitriles thus appears to be the least exothermic reversible electron attachment reaction in *n*-hexane known to-date. Actually, the standard heat of electron attachment to succinonitrile and CH2N (ca. –0.68 eV) is only slightly more negative than the standard heat of electron attachment to the ethanol tetramer cluster (ca. –0.57 eV); [15] for aceto-, butyro-, and adiponitriles, the electron attachment is less exothermic than that for the electron *solvated* by

16.

alcohol clusters in *n*-hexane! Such weakly exothermic electron attachment reactions have been observed previously only in *polar* media, where the energy of electron solvation is large. Consider, for instance, the attachment of solvated electron in tetrahydrofuran to benzene, for which the standard heat is –0.27 eV and the standard entropy is ca. –100 J/mol.K. [40]

Using a pulse radiolysis - time resolved conductivity method, Holroyd [39] obtained thermodynamic parameters for many electron attachment reactions that involve aromatic and olefinic electron acceptors in alkane liquids. All such reactions have large negative standard entropy, between –85 and –200 J/mol.K. For example, the standard entropies of electron attachment to styrene and $\alpha$–methylstyrene in hexane are –186 and –207 J/mol.K, respectively. For dinitriles the standard entropy of reaction (2) is at the lower end of this range; for mononitriles the standard entropies are unprecedentedly low. To our knowledge, electron attachment reactions with such extremely low standard entropy have never been observed, in any liquid. This suggests that electron trapping by the mononitriles occurs in a different fashion from the dinitriles and other solutes that reversibly attach the electron in *n*-hexane: whereas the dinitriles form a molecular anion, the mononitriles undergo dipole binding to the solvated electron, as was first suggested in refs. 15 and 16. The reason for such a divided pattern is examined below, using density functional theory models for the monomer and multimer radical anions.

## 4. Discussion.

### 4.1. Computational details.

The basic computational approach used in this study is similar to that used in our previous publications, in particular, ref. 4. Gas phase monomer and cluster anions were examined using DFT models with BLYP functional (Becke's exchange functional [41a] and the correlation functional of Lee, Yang, and Parr) [41b] from Gaussian 98. [42] This functional is most frequently used to estimate the energetics and hyperfine coupling constants (hfcc) in radicals and radical ions, for which it typically yields reliable results. We will use the hfcc's instead of the Mulliken spin and charge densities because the former can be experimentally determined using EPR and related spectroscopies, if only in principle. The isotropic part of the hfcc (*A*) is proportional to the *s*-character of the singly occupied molecular orbital (SOMO) for a given magnetic nucleus; the anisotropic part (only the largest principal value *2T* of the anisotropic tensor is reported) is proportional to the magnetic dipole coupling of the unpaired electron to the spin of the magnetic nucleus. [43] In the point dipole approximation, [43] for a proton at a distance *r* from the electron at the center of a spherical cavity, $2T[G] \approx 56.7/r^3[\text{Å}]$. The hfcc's for $^1$H, $^{13}$C, and $^{14}$N nuclei are given in units of G (1 Gauss = $10^{-4}$ Tesla).

A 6-31G split-valence double-ζ Gaussian basis set augmented with diffuse and polarized functions was used for all calculations (6-31+G**). A ghost chlorine atom (i.e., a set of floating-center basis functions) at the center or a cluster was added. We also carried out second-order MØller-Plesset perturbation theory [41c] calculations using the same basis set and obtained comparable results to those obtained using the DFT methods (as was also the case with ammonia clusters). We emphasize that our star-shaped model cluster anions may not resemble at all the species observed in the gas phase. [4] The species of interest to us is the *core* of a much larger cluster anion which traps the electron in its interior, or the "solvated electron" in the bulk liquid. We focus only on those aspects of these DFT models of cluster anions that are likely to relate to the observed properties of the "solvated electron" in the liquid. For the same reason, we did not focus on the energetics of such gas phase clusters. Our scope is limited to only the *structural* properties of the core anion.

Since the definition of what constitutes the cavity in a many-electron model of "solvated electron" is ambiguous, the partition of the spin density between the cavity and the solvent molecules cannot be characterized in a unique fashion. Examination of isodensity maps for SOMO suggests that the electron wavefunction inside the cavity and in *C 2p* orbitals of carbon atoms in methyl and methylene groups has opposite signs, which makes it easy to distinguish these two contributions. Qualitatively, this partition can be assessed by examination of isodensity contour maps of SOMO (like those shown below in sections 4.3 and 4.4, Figures 9 to 12). Typically, the diffuse, positive part of SOMO occupies 80-90% of the geometrical cavity at the density of +(0.01-0.12) e Å$^{-3}$ and less than 10% at the density of +0.02 e Å$^{-3}$.

### 4.2. Molecular anions of mono- and dinitriles.

The only monomer anion of a nitrile that has been thoroughly characterized by EPR is that of acetonitrile in *β–acetonitrile-d$_3$*.[31,33] Large isotropic hfcc on $^{14}$N and methyl and cyanide $^{13}$C (ca. 8, 88, and 61. 4 G, respectively) suggest considerable bending of the *C-C-N* fragment (130-145°). This angle is similar to the *H-C-N* angle in the HCN$^-$ radical anion (131°).[44] Like all other nitriles, a neutral acetonitrile molecule has the *C-C-N* angle of 180°. Bending the *C-C-N* fragment is very costly energetically, and the adiabatic electron affinity of the molecule increases as this bending decreases. It follows from the experiments and high-level *ab initio* calculations[44] that electron attachment to gas phase mononitrile molecules results in the formation of metastable dipole-bound anions that are very different structurally from the bent molecular anions observed in solid matrices, such as CH$_3$CN$^-$ and HCN$^-$. In these gas phase monoanions, the electron is electrostatically bound to the *CN* dipole, with a diffuse wavefunction that has the spatial extent of several nanometers. Since we are not interested in such dipole-bound anions, a more useful model of the nitrile anion in a liquid is provided by a DFT calculation with a

19.

basis set that is deficient in very diffuse orbitals, such as the 6-31+G** set that was used in our calculations. The validity of such an approach is suggested by the ability of small-basis DFT calculations to capture all observable properties of monomer and dimer anion of acetonitrile in crystalline *β−* and *α−* acetonitrile, respectively. [33] The resulting optimized structures are shown in Figures 6, 7, and 8; the structural parameters and energetics are summarized in Table 2, and calculated magnetic parameters (isotropic and anisotropic hfcc's) are given in Table 1S in the Supplement. The adiabatic electron affinity (EA) given in the fourth column of Table 2 was defined as the difference in the energies between the lowest neutral and negatively charged nitrile molecules *in the same conformation*. For all dinitriles, the *trans-* conformation in which the two CN dipoles are opposing each other is always the lowest energy one. The *C-C-N* angle strongly deviates from 180º for all (di-) nitrile anions (Table 2), the *C-N* bond is elongated from 1.174 Å to 1.18-1.21 Å, and the *C-CN* bond is elongated from 1.47 Å to 1.49-1.52 Å. The degree of bending and bond elongation decreases with the size of the molecule, because these correlate with the degree of sharing of the negative charge with the alkyl/methylene groups. Figures 6, 7, and 8 show the isodensity contour maps of the SOMO; it is seen from these maps and Table 2S that a large fraction of the spin density is contained in the *N* and *C 2p* orbitals of *CN* groups. The larger the bending angle, the more of the spin density is in these groups; for large dinitriles (and polynitriles considered in section 4.5), the electron density spreads over, and the bending decreases.

For all of these (di-) nitriles, the electron affinity as defined above is strongly negative (Table 2); the most negative is the EA for mononitriles (ca. -1 eV), the least negative are the EA for malono- and succino- nitriles (-0.27 and -0.34 eV, respectively; Table 2). In a nonpolar liquid, such as *n*-hexane, the free energy for solvation of an anion is mainly determined by the Born polarization energy. [8] A better estimate of this solvaton

20.

energy can be obtained using the polarized continuum method of Tomasi et al;[46] this method was used in the integral equation formulation (IEFPCM). The solvation energy for the anions and corresponding neutral molecules (for their optimized gas-phase geometries) are given in the fifth and sixth columns of Table 2, and the last column gives $EA_{liq}$, which is defined as the adiabatic affinity of a solvated nitrile molecule to the electron in vacuum. Observe that all of these estimates are either very close to zero (for mononitriles) or positive (for dinitriles). Given that the $V_0$ (the energy of $e_{qf}^-$ relative to the vacuum) for *n*-hexane is slightly positive (ca. 100 meV at 295 K)[8] and that $e_{solv}^-$ is ca. 180-200 meV [8,9] lower in energy than $e_{qf}^-$, the adiabatic affinity of a solvated nitrile molecule to a quasifree/solvated electron is, respectively, 0.1 eV greater or 0.1 eV lower than $EA_{liq}$. It is seen from Table 2 that the largest absolute standard heat for reaction (2) is predicted for succinonitrile and malononitrile, in full agreement with the experiment. In particular, for malononitrile, the heat of electron attachment is so negative that the reaction would be irreversible on our time scale, as is indeed the case (section 3). The attachment is weakly exothermic for other dinitriles, but it is weakly endothermic for the mononitriles. Thus, the DFT calculations rationalize the pattern observed in section 3: for the dinitriles, the electron attachment with the formation of molecular anions with bent *C-C-N* fragments is sufficiently exothermic for reaction (2) to occur (Table 1). For the mononitriles, this type of electron stabilization is energetically prohibitive, and some other mechanism for electron localization sets in. We turn now to the examination of this mechanism.

### 4.3. "Solvated electron" in alkanes.

One-electron cavity models for electron solvation in *n*-alkanes are much less developed and detailed than such models for electrons in water, ammonia, and other polar liquids. We have recently reviewed these models in section V.B of ref. 9, so only a brief

21.

outline is given here. As is the case with polar liquids, the classical potential is prescribed for the electron, by introducing a matrix of *C-C* and *C-H* bond polarizabilities [47] (the most recent and advanced of such models is given by Quirke and co-workers). [48] With suitable parameters, it is possible to qualitatively reproduce the formation of an electron trapping cavity in *n*-alkanes other than methane and in polyethylene and account for, *inter alia*, the pressure and temperature dependence of quasifree electron energy vs. vacuum, $V_0$. [47,48] From the practical standpoint, however, it appears that all of the observed properties of solvated electron in liquid alkanes are already captured in the simplest "electron bubble" model in which the electron resides inside a spherical potential well with a hard radius *a* of 3.4-3.7 Å. [9,12] The ground state electron has a spherical *s*-wavefunction and a binding energy of 180-200 meV; when this electron absorbs light, it is promoted directly into the conduction band of the liquid (becoming a *p*-wave). This "electron bubble" model consistently accounts for the absorption spectra of solvated and trapped electrons in liquid [9] and vitreous alkanes, [12] respectively, magnetic resonance data, [10] reaction volumes for electron attachment, [8] and the recent terahertz spectroscopy data. [9,49] What these results suggest is that the specific origin of the electron-trapping potential might not be particularly important for understanding the observed properties of this species. What matters more is a realistic estimate for the cavity size and binding energy, as these two parameters determine the spatial extent of the wavefunction that in turn defines these observable properties. Conversely, little about the exact origin of the trapping potential can be surmised directly from the available experimental data.

The same difficulty exists for some other "solvated electron" species, such as ammoniated electron, for which there are experimental results that are difficult, if not impossible, to account for in the standard one-electron model. [4,7] In the previous study, we demonstrated how magnetic resonance data (specifically, Knight shifts for $^{14}$N and $^1$H nuclei) [6] can be consistently accounted for in the DFT radical anion model of

22.

ammoniated electron.[4] The latter model is a further development of the qualitative model suggested by Symons[7] and *ab initio* calculations by other authors.[5] In this type of model, the solvation complex is regarded as a multimer radical anion in which the electron density is divided (mainly) between several nitrogen atoms, residing in their frontier *N 2p* orbitals. Whereas the standard cavity model cannot account for the large spin density in the *N 2s* orbitals of ammonia molecules (i.e., large positive hfcc's for $^{14}$N nuclei) and small negative hfcc's for $^1$H nuclei (attained via spin-bond polarization involving *N 2p* orbitals),[4] the radical anion model provides realistic estimates for these hfcc parameters and naturally accounts for the observed pattern. It also qualitatively accounts for the small downshift in the *N-H* stretching and *H-N-H* bending modes for ammonia molecules that are involved in the electron solvation[50] (analogous to those observed by time-resolved Raman spectroscopy for hydrated electron).[51]

Below, we use a similar approach to examine electron solvation by alkanes, acetonitrile, and mixed alkane-acetonitrile clusters. Due to the computational limitations, the alkane of choice is ethane, as it is the simplest hydrocarbon liquid for which localized (as opposed to quasifree) electrons are observed.[8,47,48] The same considerations limited the size of the model cluster, which consisted of five molecules arranged on a triangular dipyramid pattern with approximate $C_{3h}$ symmetry (Figure 9). This star shaped arrangement allows exact definition of the "cavity" and its center. It is noteworthy that magnetic resonance studies of Kevan et al.[10] suggested that the number of magnetically equivalent methyl protons that are dipole coupled to the electron is ca. 20, so the coordination number of the electron is comparable to the one in this model ($\approx$7 methyl groups?). As shown below, there are reasons to believe that this coordination number might be lower, because the point-dipole approximation used in ref. 10a would break down in the radical anion picture of the "solvated electron."

23.

We first consider the pentamer radical anion of ethane. As seen from the SOMO maps shown in Figure 9, in this anion the spin density resides mainly in the *C 2p* orbitals with the long axis along the *C-C* bonds. The cavity is filled by a diffuse electron wavefunction; this cavity has slight axial elongation (the summary of geometry and magnetic parameters is given in Table 3). Most of the electron density, however, is in the carbon orbitals. The cavity itself is formed by Coulomb repulsion between ethane molecules (each one of which carries a fractional negative charge) rather than by Pauli exclusion of closed-shell ethane molecules. Since the negative charge is spread over several molecules, the changes in the bond lengths and bond angles as compared to neutral molecules are very slight: e.g., the *C-C* bonds are elongated from 1.5435 Å to 1.544 Å, the *C-H* bonds are elongated from 1.103 Å to 1.104 Å, and the *C-C-H* angle increased from 111.4º to 111.6º (for the inner methyl groups). The spin density is unevenly divided between the inner and outer carbon atoms. Mulliken population analysis yields 0.166 for $C_a$ (Figure 9) and 0.076 for $C_b$, and isotropic hfcc's for $^{13}$C nuclei follow the same pattern: the inner carbons have large hyperfine constants (18.4 G and 20.6 G, for in-plane and axial carbons, respectively), whereas hfcc's for the outer carbons are 4 times smaller (4.6 and 5.4 G, respectively).

These large isotropic hfcc's for methyl $^{13}$C nuclei are the key prediction of the radical anion model. Unfortunately, no such EPR data for trapped electrons in vitreous hydrocarbons are currently available, but this prediction provides the means to falsify the radical anion model and discriminate between this model and the cavity model.

The hydrogen atoms in the methyl groups forming the cavity are 3.95 Å away from the center; the second shell of the hydrogen atoms (in the outer methyl groups) is 6.25 Å from the center. Holroyd [8] gives 3.2-3.6 Å for the cavity radius, and Kevan et al. [10] gives 3.4-3.6 Å. It should be stressed that electron spin echo envelope modulation (ESEEM) spectroscopy used by Kevan et al. [10a] gives only the total dipolar field on the

24.

protons.[43] The size of the cavity is estimated by assuming that $N$ (magnetically-equivalent) protons at a distance $a$ from the cavity center interact, via dipole-dipole interaction, with the point electron at the cavity center. Thus, the ESEEM spectrum only suggests that that total sum of anisotropic hfcc's (which is proportional to $N/a^3$) is ca. 22-26 G. Figure 9 suggests that this point dipole approximation would not hold in the radical anion model because the spin density is spread over the alkane molecules. Consequently, the $H_a$ protons have $2T \approx 0.64$-$0.78$ G (vs. 0.92 G in the point-dipole approximation), whereas $H_b$ protons have $2T \approx 0.32$-$0.36$ (vs. 0.23 G in the same approximation; see Table 3). The sum of these constants is already 16 G, despite the fact that the model cavity is much larger than assumed by Kevan et al.[10] Since more (and longer chain) alkane molecules are involved in the electron trapping in vitreous alkanes, the cavity should be smaller than estimated in our gas-phase calculations (since Coulomb repulsion is reduced by screening and spreading of the charge. Thus, the dipolar field of the protons observed by Kevan et al.[10] naturally emerges from the radical anion model. Interestingly, this model predicts small, negative isotropic hfcc's on the protons (for the ethane pentamer anion in Figure 9, these are –0.3 to -0.7 G); potentially, this prediction might also be observable experimentally.

What is the expected change for longer alkane molecules? The answer can be surmised by replacing in-plane ethane molecules by, for example, $n$-butane. The hfcc's for carbon-13 nuclei starting from $C_a$ are 14.5, 5.8, 3.7, and 5.7 G; the hfcc's for the corresponding protons decrease as –0.6, -0.47, -0.2, and –0.19 G. The changes in the geometry of the butane molecules required to accommodate the excess electron are even smaller then those for the ethane molecules in the pentamer anion; for larger hydrocarbon molecules these changes would be smaller still. Thus, the reaction volume for electron localization is still determined by the size of the cavity, rather than the total volume of the electron density on alkane molecules, which explains why their reaction volumes

25.

correspond so well to the cavity sizes.[8,9] As seen from the hfcc pattern, the SOMO is spread over *C 2p* orbitals (as is the case for ethane) gradually fading out on the odd-even pattern from one carbon atom to the next. What that implies for a large cluster anion or "solvated electron" in the liquid is that the SOMO is spread along the hydrocarbon chains, reaching far out of the cavity, thereby filling the spherical layer of thickness comparable to the length of the hydrocarbon molecule. Since several such molecules protrude from the cavity, the overall distribution of the spin density is roughly spherical, as one would expect for an *s*-function. The bulk of the spin density is in the frontier orbitals of carbon atoms in methyl (or methylene) groups at the cavity wall but it also spreads out a few bonds away from the cavity and some of this density occupies the cavity and the inter-chain spaces. All of the features of the ground state function as given by the "electron bubble" model are thus recovered. In terms of the energetics, spreading of the electron density, regardless of its exact origin, is equivalent to lowering the binding energy. The resulting picture of electron localization in liquid alkanes is not dissimilar to that obtained using local density approximation Car-Parrinello calculations [52a] for crystalline polyethylene, where self-trapped electron polaron was shown to spread along 6-7 methylene units, by occupying both the interchain spaces and *C 2p* orbitals of a single chain. Calculations of Serra et al. [52b] suggest that the balance between this on- and inter-chain electron density is precisely what determines the electron conduction in polyethylene.

One of the advantages of this radical anion picture of the "solvated electron" in liquid alkanes is the ease with which solute molecules can be incorporated in the solvent network.

### 4.4. Mononitriles.

Liquid acetonitrile solvates the electron in a way similar to that observed for alkanes (exhibiting a broad absorption band in the near infrared), and it is reasonable to expect that the electron is localized in a similar fashion. The DFT calculation for a star shaped pentamer acetonitrile anion shown in Figure 10 validates this expectation. The spin density is divided mainly between the carbon atoms of methyl groups forming the cavity; some of this spin density fills the cavity. The average center of mass – hydrogen distance is 3.55 Å, but the SOMO is much more localized than the SOMO for the ethane (Figure 9). The changes in the molecular geometry as compared to neutral, isolated acetonitrile molecule are minor (the bond lengths are only 0.1-0.2% longer); see Table 3. The *C-C-N* angle is within $0.27°$ from $180°$, as opposed to the monomer and dimer anions of acetonitrile (for which this angle is ca. $130°$). This small bending explains the relative stability of the star-shaped solvation complex. The isotropic hfcc on methyl carbon-13 is 26.7 and 35 G (for axial and planar molecules); the hfcc's for other nuclei are significantly lower (Table 3).

In Figure 11 a mixed cluster anion is shown in which acetonitrile substitutes for one of the ethane molecules. This substitution results in a surprisingly small change in the structure of the anion as compared to the ones shown in Figures 9 and 10. The *C-C-N* angle in the acetonitrile molecule is still very close to $180°$ (within $0.3°$), and the changes in the geometry are minor (in a stark contrast to the acetonitrile monomer). The spin density is unevenly divided between the two carbons in the ethane molecules and the methyl carbon of acetonitrile. The isotropic hfcc constants on $^{13}$C nuclei are 55.4 G (for $^{13}$CH$_3$CN), 15-18 G (for $C_a$), and 1.5-3 G (for $C_b$), i.e. the spin density stays mainly at the cavity wall. The *CN* dipole is ca. 6.4 Å away from the cavity center (as compared to 6 Å estimated in ref. 15 using the electron bubble model). The geometry and magnetic parameters for this cluster are given in Table 3. The conclusion is that with very minor geometry change, the acetonitrile molecule can participate in the network of alkane molecules that share the charge and form a multimer radical anion. The structure shown in Figure 11 is a radical anion, just like the structures shown in Figures 6, 7, and 8, but the mechanism for the stabilization of this anion is different: the solute molecule shares the negative charge with the solvent molecules, undergoing almost no change in the geometry. In molecular anions shown in, e.g., Figure 8, the solute molecule accepts the whole charge on itself, which requires an extensive change in the geometry of the acceptor. For mononitriles, the first mode of electron localization prevails, hence the very small standard heat and entropy of reaction (2). For dinitriles, the molecular anion is formed instead, hence the large standard heat and entropy of reaction (2). However, the comparison of large dinitrile anions (CH2N) in Figure 8c and the composite cluster anion in Figure 11 suggests a remarkable degree of structural similarity. In the former anion, the *C-C-N* angle is only 12º smaller than 180º and consequently it shares many common features with the multimer anion shown in Figure 11. In both of these cases, the spin density is mainly in *C 2p* orbitals on the alkane chains. The specific way in which these spin densities are shared between the carbon atoms is different, because the aliphatic scaffolding is different, but the general pattern of the SOMO is the same. These two patterns become even more similar for the polynitrile anions examined next.

**4.5. Polynitrile rings.**

When several nitrile groups are placed on a cycloalkane ring as shown in Figures 1c and 2S(a,b), the *C-C-N* angle becomes even closer to 180º (due to extensive negative charge sharing by the several nitrile groups) and the overall change in the geometry after the electron attachment becomes even smaller than that for the CH2N anion. Figure 12 shows isodensity SOMO maps for two polynitrile anions (whose structures were

28.

geometry optimized using the BLYP/6-31+G** method): 1,3,5,7-tetracyano-cyclooctane anion (CO4N$^-$) and 1,2,4,5,7,8,10,11-octacyano-cyclododecane anion (CD8N$^-$). More detailed maps are given in Figures 7S, 8S, and 9S in the Supplement, and the isotropic hfcc's are summarized in Figure 10S. The lowest energy CO4N$^-$ anion has $C_i$ symmetry, and the CD8N$^-$ anion has $C_2$ symmetry (a sligtly distorted $C_4$ structure). In the CO4N$^-$ anion, the *C-C-N* angles are 168.5° and 172°; for the CD8N$^-$ anion, all *C-C-N* angles are within 6° of 180°. Whereas relatively large *C-C-N* bending angles and small ring size preclude the formation of a "cavity electron" for CH2N$^-$ and CO4N$^-$ anions (Figures 12(a) and 7S), the SOMO maps for CD8N$^-$ reveal a dumbbell shaped diffuse electron orbital filling the cavity (Figures 12(b), 8S, and 9S). Although most of the spin density still resides in the *C-C-N* fragments (chiefly, on the carbons in the cyclododecane ring), as is the case with the smaller rings, the partitioning of the spin density between the cavity and the cyanide fragments is not too different from that observed for multimer radical anions shown in Figures 10 and 11. The orbitals on these *C-C-N* fragments resemble the orbitals for acetonitrile molecules in Figures 10 and 11, though in CD8N$^-$ the SOMO exhibits more distinctive *N 2p* character (due to the bending of the *C-C-N* fragment). The electron cavity is formed by eight C-H groups whose protons alternate above and below the midplane of the ring (Figures 12(b) and 10S). The distance between the opposing >CH hydrogens is ca. 2.8 Å, i.e., the electron cavity is tight (for acetonitrile, the cavity radius was estimated to be ca. 3.55 Å). Consequently the isotropic hfcc's on the protons are positive (ca. +1.2 G) despite significant bond spin polarization involving *C 2p* orbitals. The isotropic hfcc's for cyanide $^{13}$C and $^{14}$N nuclei are small and positive, +0.43 G and +0.68 G, respectively (vs. +1.5 G and +1.2 G, respectively, for the acetonitrile hexamer anion in Figure 10). The total hfcc for $^{13}$C nuclei in the cyclododecane ring is ca. +110 G, whereas for methyl carbons in the hexamer anion shown in Figure 10 this parameter is ca. +160 G. Paradoxically, CD8N$^-$ more closely resembles the idealized "cavity electron"

29.

postulated by one-electron models than the solvated electron occurring in neat acetonitrile.

In summary, in the multimer radical anion picture of the "solvated electron," CH2N$^-$ and the two polynitrile anions shown in Figure 12 might be viewed as the sought "encapsulated electron" because their general electronic structure is very similar to the structure of the multimer anions shown in Figures 9, 10, and 11. For the CD8N$^-$ anion shown in Figures 12(b), there is a "cavity electron" (Figures 8S(a,b) and 9S(a,b)) confined inside the ring. This "cavity electron" is stabilized by eight CN dipoles pointing towards the cavity center. While the intracavity orbital carries only a fraction of the total spin density, such is also the case with the "solvated electron" occurring in neat alkane and acetonitrile solvents. A sharp division between these two classes of anions does not appear to exist: The "encapsulated electron" is a large molecular anion whose electronic structure strikingly resembles a radical anion commonly identified as the "solvated electron." Apart from this difference in terminology, there might be little difference in the nature of the species involved.

## 5. Conclusion.

In this paper, we attempt to expand the concept of "solvated electron" to supramolecular structures. Two general approaches to electron encapsulation by a neutral molecule are recognized: a cage that is lined by polar groups on the inside of the cavity and a cage that has polar groups arranged outside of the cavity that dipole bind the electron. The latter approach makes it easier to design such a cage using organosynthetic methods, but it is feasible for polynitrile cages only. That prompted us to study electron attachment to mono- and di- nitriles molecules dispersed in *n*-hexane, which has very low electron binding energy, thus allowing electron attachment to the solute. These studies indicated that electron trapping by mono- and di- nitriles occurs in two different ways, as suggested by the thermodynamics of reaction (2): the dinitriles form a molecular anion, whereas mononitriles substitute for a solvent molecule at the cavity of the "solvated electron." We then proceeded to model these two modes of electron attachment using DFT method. It was realized that the two ways of electron localization stem from the high energetic cost of bending the *C-C-N* fragment. This allows the formation of a molecular anion for dinitriles, where this bending is reduced due to the charge sharing between two nitrile groups, but prohibits it for mononitriles, for which such a sharing cannot occur. As a result, the charge is shared instead with the solvent in a kind of extended radical anion analogous to the "solvated electron." This radical anion picture can also be applied to the "solvated electron" itself (in liquid alkanes), and it seems to produce a species whose SOMO exhibits all of the desirable features of the ground state wave function given by the electron bubble model that successfully rationalizes experimental observations. This refinement of the standard "solvated electron" model makes new predictions as to the electron structure of the "solvated/trapped electron" in hydrocarbons. Specifically, we predict that there are large isotropic hyperfine coupling constants for $^{13}$C nuclei in the methyl/methylene groups at the cavity wall due to the substantial transfer of the spin density into the frontier *C 2p* orbitals of these groups (similar to the situation for the ammoniated electron). [4] We suggest that the spin density spreads considerably (over a few *C-C* bonds) along the aliphatic chain. Such a view appears to be compatible with the available $^1$H ESEEM data for trapped electrons in vitreous alkanes. [10a] Since EPR parameters for trapped electrons in $^{13}$C labeled hydrocarbons are currently unavailable, this key prediction of the model cannot presently be verified, but it is a novel, verifiable prediction that can settle the matter in an unambiguous fashion.

The radical anion picture of the "solvated electron" in hydrocarbons readily rationalizes substitution at the cavity wall in the presence of mononitriles. Due to the extensive charge sharing between the molecules, very little change in the geometry of the solvent/solute molecules or the cavity size is needed to accommodate a nitrile molecule. The comparison of spin density maps for the mixed cluster anion shown in Figure 11 and large di- (e.g., Figure 8c) and poly- nitrile ring anions (Figure 12) suggests striking similarity between the ways in which the spin density is shared (mainly) by *C 2p* orbitals in both of these cases. If the central cavity in the polynitrile trap is sufficiently large and the *C-C-N* bending angle is reduced due to the extensive negative charge sharing between several nitrile groups, a feature resembling of the "cavity electron" (as observed in Figures 9 and 10) emerges (Figure 12b) It appears that there is no conceptual difference between certain types of "solvated electrons" and "molecular anions" as these two are different realizations of the same motif for electron localization. In this sense, electron encapsulation by supramolecular structures is possible. One the other hand, the task of finding a supramolecular structure that experimentally implements (or closely approximates) the highly idealized "cavity electron" favored by one-electron models might be impossible. For polynitrile electron traps, this task is certainly impossible, as suggested by the results of the present study; for hydroxyl lined cages (Figure 1b) discussed in section 1.1, the matter is still unsettled.

Apart from a purely academic interest, encapsulated electrons might present a technological one. Indeed, such a species would be the organic chemistry analog of a semiconductor 3D quantum well or a quantum dot holding a single charge. Since trapped electron is largely decoupled from the magnetic nuclei in the solvent cage (such as the protons), the spin relaxation times are typically long, ranging from tens of milliseconds at 4 K to microseconds at 300 K. By contrast, decoherence times for electrons in semiconductor quantum dots are often shorter than 1 µs, even at 1-4 K. Furthermore, semiconductor dots/wells have multiple defects and exhibit a wide distribution of shapes and sizes that often preclude their reliable use in the circuitry; by contrast, electron trapping cages would be highly homogeneous in their properties, easy to manipulate in the solution, and easily modified and linked to other molecular structures.

## 6. Acknowledgement.

IAS thanks C. D. Jonah, R. A. Holroyd, J. F. Wishart, and F. T. Williams for many useful discussions and F.-P. Schmidchen, J.-M. Lehn, J. L. Dye, J. Sessler, and E. Anslyn for their advice and assistance. The initial impetus for this study was provided by a discussion with J. R. Miller that happened nearly ten years ago. This work was supported by the Office of Science, Division of Chemical Sciences, US-DOE under contract number W-31-109-ENG-38.

*Supporting Information Available:* A single PDF file containing (1.) Appendix: Details of Kinetic Analysis (2.) Table 1S; (3.) Figs. 1S to 10S with captions. This material is available free of charge via the Internet at http://pubs.acs.org.

## References.

(1) Shkrob, I. A.; Sauer, M. C., Jr. In *Charged Particle and Photon Interactions with Matter*, Mozumder, A.; Hatano, Y., Eds.; Marcel Dekker: New York, 2004; pp. 301.

(2) Hart, E. J.; Anbar, M. *The Hydrated Electron*, Wiley-Interscience: New York, 1970; Coe, J. V. *Int. Rev. Phys. Chem.* **2001**, *20*, 33; Kevan, L. *Adv. Radiat. Chem.* **1974**, *4*, 181.

(3) Rossky, P. J.; Schnitker, J. *J. Phys. Chem.* **1988**, *92*, 4277; Schnitker, J.; Motakabbir, K.; Rossky, P. J.; Friesner, R. A. *Phys. Rev. Lett.* **1988**, *60*, 456; Webster, F. J.; Schnitker, J.; Frierichs, M. S.; Friesner, R. A.; Rossky, P. J. *Phys. Rev. Lett.* **1991**, *66*, 3172; Webster, F. J.; Rossky, P. J.; Friesner, R. A. *Comp. Phys. Comm.* **1991**, *63*, 494; Motakabbir, K.; Schnitker, J.; Rossky, P. J. *J. Chem. Phys.* **1992**, *97*, 2055; Rosenthal, S. J.; Schwartz, B. J.; Rossky, P. J. *Chem. Phys. Lett.* **1994**, *229*, 443; Murphrey, T. H.; Rossky, P. J. *J. Chem. Phys.* **1993**, *99*, 515; Schwartz, B. J.; Rossky, P. J. *J. Chem. Phys.* **1994**, *101*, 6917; *J. Phys. Chem.* **1994**, *98*, 4489; *Phys. Rev. Lett.* **1994**, *72*, 3282; *J. Chem. Phys.* **1994**, *101*, 6902; Wong, K. F.; Rossky, P. J. *J. Phys. Chem. A* **2001**, *105*, 2546; Borgis, D.; Staib, A. *Chem. Phys. Lett.* **1994**, *230*, 405; Staib, A.; Borgis, D. *J. Chem. Phys.* **1995**, *1995*, 2642; Borgis, D.; Staib, A. *J. Chim. Phys.* **1996**, *39*, 1628; *J. Chem. Phys.* **1996**, *104*, 4776; *J. Phys.: Condens. Matter* **1996**, *8*, 9389; Staib, A.; Borgis, D. *J. Chem. Phys.* **1996**, *104*, 9027; Borgis, D.; Bratos, S. *J. Mol.Struct.* **1997**, *1997*, 537; Nicolas, C.; Boutin, A.; Levy, B.; Borgis, D. *J. Chem. Phys.* **2003**, *118*, 9689. (b) Boero, M.; Parrinello, M.; Terakura, K.; Ikeshoji, T.; Liew, C. C. *Phys. Rev. Lett.* **2003**, *90*, 226403.

(4) Shkrob I. A., *J. Phys. Chem. A*, submitted 2005 (ms# jp050564v); preprint available on on http://www.arXiv.org/abs/physics/0509137.

(5) Newton, M. D. *J. Phys. Chem.* **1975**, *79*, 2795; Clark, T.; Illing, G. *J. Am. Chem. Soc.* **1987**, *109*, 1013.

(6) Catterall, R.; Stodulski, L. P.; Symons, M. C. R. In *Metal-Ammonia Solutions; Collogue Weyl II*; Lagowski, J. J., Sienko, M. J., Eds.; Butterworths: London, 1969; pp 151 and references therein; Niibe, M.; Nakamura, Y. *J. Phys. Chem.* **1984**, *88*, 5608; O'Reilly, D. E. *J. Chem. Phys.* **1964**, *41*, 3729; Holton, D. E.;

Edwards, P. P.; McFarlane, W.; Wood, B. *J. Am. Chem. Soc.* **1983**, *105*, 2104; Catterall, R.; Stodulski, L. P.; Symons, M. C. R. *J. Chem. Society A:* **1968**, 437.

(7) Symons, M. C. R. *Chem. Soc. Rev.* **1976**, *5*, 337.

(8) Holroyd, R. A. in *Charged Particle and Photon Interactions with Matter*, Mozumder, A.; Hatano, Y., Eds.; Marcel Dekker: New York, 2004; pp. 175.

(9) Shkrob, I. A.; Sauer, Jr., M. C. *J. Chem. Phys.* **2005**, *122*, 134503.

(10) (a) Feng, D.-F.; Kevan, L.; Yoshida, Y. *J. Chem. Phys.* **1974**, *61*, 4440; (b) Narayana P. A.; Kevan, L. *J. Chem. Phys.* **1976**, *65*, 3379.

(11) K. Hiraoka, J. Phys. Chem. **85**, 4008 (1981); T. Kimura, K. Fueki, P. A. Narayana, and L. Kevan, Can. J. Chem. **55**, 1940 (1977); M. Nishida, J. Chem. Phys. **65**, 242 (1977).

(12) McGrane, S. D.; Lipsky, S. *J. Phys. Chem. A* **105**, *2001*, 2384; Funabashi, K. *Adv. Radiat. Chem.* **1974**, *4*, 103; Ichikawa, T.; Yoshida, Y. *J. Chem. Phys.* **1981**, *75*, 5432.

(13) Hammer, H.; Schoepe, W.; Weber, D. *J. Chem. Phys.* **1976**, *64*, 1253; T. Ichikawa, T.; Yoshida, H. *J. Chem. Phys.* **1981**, *75*, 5432.

(14) Mozumder, A. *J. Phys. Chem.* **1972**, *76*, 3824; Kenney-Wallace, G. A.; Jonah, C. D. *J. Phys. Chem.* **1982**, *86*, 2572; Baxendale, J. H. *Can. J. Chem.* **1977**, *55*, 1996; Baxendale, J. H.; Sharpe, P. H. G. *Chem. Phys. Lett.* **1976**, *41*, 440; Ahmad, M. S.; Atherton, S. J.; Baxendale, J. H. Radiat. in Radiat. Res., Proc. Int. Congr. 6th, 1979, pp. 220; Baxendale, J. H.; Rasburn, E. J. *J. Chem. Soc. Farad. Trans. 1* **1974**, *70*, 705; Baxendale, J. H.; Keene, J. P.; Rasburn, E. J. *J. Chem. Soc. Farad. Trans. 1* **1974**, *70*, 718; Gangwer, T. E.; Allen, A. O.; Holroyd, R. A. *J. Phys. Chem.* **1977**, *81*, 1469; Smirnov, S. N.; Anisimov, O. A.; Molin, Y. N. *Chem. Phys.* **1986**, *109*, 321.

(15) Shkrob, I. A.; Sauer, Jr., M. C. *J. Phys. Chem. A* **2005**, *109*, 5754.

(16) Shkrob, I. A.; Sauer, Jr., M. C. *J. Phys. Chem. A* **2002**, *106*, 9120.

(17) Lund, A. *Res. Chem. Intermed.* **1989**, *11*, 37 and references therein.

(18) Dye, J. L. *Ann. Rev. Phys. Chem.* **1987**, *38*, 271 and references therein.

(19) Rao, K. V. S.; Symons, M. C. R. *J. Chem. Soc., Farad. Trans. 2* **1972**, *68*, 2081; Brooks, J. M.; Dewald, R. R. *J. Phys. Chem.* **1968**, *72*, 2655; Mal'tsev, E. I.; Vannikov, A. V. *High Energy Chem.* **1971**, *5*, 337; *High Energy Chem.* **1974**, *7*, 338.

(20) Dietrich, B. *Pure & Appl. Chem.* **1993**, *65*, 1457.

- Bianchi, A.; Bowman-James, K.; Garcia-Espana E. *Supramolecular Chemistry of Anions*, John Wiley – VCH: New York, 1997.

(22) Smith, P. H.; Barr, M. E.; Brainard, J. R.; Fird, D. K.; Freiser, H.; Muralidharan, S.; Reilly, S. D.; Ryan, R. R.; Silks, III, L. A.; Yu, W. *J. Org. Chem.* **1993**, *58*, 7939; *Inorg. Chem.* **1995**, *34*, 569.

(23) Wishart, J. F.; Neta, P. J. Phys. Chem. B. 2003, 107, 7261.

(24) Beer, P. D.; Gale, P. A. *Angew. Chem. Int. Ed.* **2001**, *40*, 486 and references therein.

(25) Schreerder, J.; Engbersen, J. F. J.; Reinhoudt, D. N. *Recl. Trav. Chim. Pays-Bas.* **1996**, *115*, 307.

(26) Xia, C.; Peon, J.; Kohler, B. *J. Chem. Phys.* **2002**, *117*, 8855.

(27) Mitsui, M.; Ando, N.; Kokubo, S.; Nakajima, A.; Kaya, K. *Phys. Rev. Lett.* **2003**, *91, 153002*

(28) (a) Takayanagi, T. *Chem. Phys. Lett.* **2004**, *302*, 85; (b) *J. Chem. Phys.* **2005**, *122*, 244307.

(29) Shaede, E. A.; Dorfman, L. M.; Flynn, G. J.; Walker, D. C. *Can. J. Chem.* **1973**, *51*, 3905.

(30) Sprague, E. D. *Ph. D. Thesis*, University of Tennessee, June 1971, pp. 115-117.

(31) Williams F.; Sprague, E. D. *Acc. Chem. Res.* **1982**, *15*, 408; Bonin, M. A.; Chung, Y. J.; Sprague, E. D.; Takeda, K.; Wang, J. T.; Williams, F. *Nobel Symp.* **1972,** *22*, 103.

(32) (a) Bonin, M. A.; Takeda, K.; Tsuji, K.; Williams, F. *Chem. Phys. Lett.* **1968**, *2*, 363; *J. Chem. Phys.* **1969**, *50*, 5423; Campion, A.; Williams, *J. Chem. Phys.*

**1971**, *54*, 4510; (b) Campion, A.; Ghormley, J. A.; Williams, F. *J. Am. Chem. Soc.* **1972**, *94*, 6301.

(33) Shkrob, I. A.; Takeda, K.; Williams, F. *J. Phys. Chem. A* **2002**, *106*, 9132.

(34) Bell, I. P.; Rodgers, M. A. J.; Burrows, H. D. *J. Chem. Soc.* **1976**, 315.

(35) Mozumder, A. *Res. Chem. Intermed.* **1999**, *25*, 243; *Chem. Phys. Lett.* **1993**, *207*, 245; *Chem. Phys. Lett.* **1995**, *233*, 167; *J. Phys. Chem.* **1996**, *100*, 5964.

(36) Stass, D. V.; Sviridenko, F. B.; Molin, Yu. N. *Radiat. Phys. Chem.* **2003**, *67*, 207.

(37) Table VII.4 & 5 in Tabata, Y.; Ito, Y.; Tagawa, S. *CRC Handbook of Radiation Chemistry*, CRC Press: Boca Raton; 1991, p. 414.

(38) Holroyd, R. A.; Nishikawa, M. *Radiat. Phys. Chem.* **2002**, *64*, 19.

(39) Holroyd, R. A. *Ber. Bunsenges. Physik. Chem.* **1977**, *81*, 298

(40) Marasas, R. A.; Iyoda, T.; Miller, J. R. *J. Phys. Chem. A* **2003**, *107*, 2033.

(41) (a) Becke, A. D. *Phys. Rev. A* **1988**, *38*, 3098; (b) Lee, C., Yang, W.; Parr, R. G. *Phys. Rev. B* **1988**, *37*, 785 ; (c) MØller, C.; Plesset, M. C. *Phys. Rev.* **1934**, *46*, 618; Frish, M. J.; Head-Gordon, M.; Pople, J. A. *Chem. Phys. Lett.* **1988**, *153*, 503, ibid. **1990**, *166*, 275; ibid. **1990**, *166*, 281.

(42) Frisch, M. J. et al, *Gaussian 98, revision A.1*, Gaussian, Inc., Pittsburgh, Pennsylvania, 1998.

(43) Dikanov, S. A.; Tsvetkov, Y. D. *Electron Spin Echo Envelope Modulation (ESEEM) Spectroscopy*; CRC Press: Boca Raton; 1992.

(44) Adrian, F. J.; Cochran, E. L.; Bowers, V. A.; Weatherley, B. C. *Phys. Rev.* **1969**, *177*, 129

(45) for example, Abdoul-Carime, H.; Bouteiller, Y.; Desfrançois, C.; Philippe, L.; Schermann, J. P. *Acta. Chem. Scand.* **1997**, *51*, 145; Desfrançois, C.; Abdoul-Carmie, H.; Adjouri, C.; Khelifa, N.; Schermann J.P *Europhys. Lett.* **1994**, *26*, 25.

(46) Tomasi, J.; Persico, M. *Chem. Rev.* **1994**, *94*, 2027; Cancès, E.; Mennucci, B.; Tomasi, J. *J. Chem. Phys.* **1997**, *107*, 3032.

(47)   Coker, D. F. *J. Chem. Phys.* **1992**, *96*, 652; Coker, D. F.; Berne, B. J. In *Excess Electrons in Dielectric Media*, Ferradini, C.; Jay-Gerin, J.-P.; Eds; CRC Press: Boca Raton, 1991; pp. 211 and references therein; Sheu, S.-Y.; Cukier, R. I. *J. Chem. Phys.* **1991**, *94*, 8258; Space, B.; Coker, D. F.; Liu, Z. H.; Berne, B. J.; Martyna, G. *J. Chem. Phys.* **1992**, *97*, 2002; Chen J.; Miller, B. N. *J. Chem. Phys.* **1994**, *100*, 3013; Chandler, D. *J. Chem. Phys.* **1990**, *93*, 5075 and references therein.

(48)   Cubero D.; Quirke, N. *J. Chem. Phys.* **2004**, *120*, 7772; Meunier, M.; Quirke, N. *J. Chem. Phys.* **2000**, *113*, 369.

(49)   Knoesel, E.; Bonn, M.; Shan, J.; Heinz, T. F. *Phys. Rev. Lett.* **2001**, *86*, 340; *J. Chem. Phys.* **2004**, *121*, 394.

(50)   DeBacker, M. G.; Rusch, P. F.; DeBettignies, B.; Lepoutre, G. In *Electrons in Fluids*; Jortner, J., Kestner, N. R., Eds.; Springer-Verlag: New York, 1973; pp 169; Smith, B. L.; Koehler, W. H. ibid., pp. 145; Rusch, P. F.; Lagowski, J. J., ibid., pp 169.

(51)   Tauber, M. J.; Mathies, R. A. *Chem. Phys. Lett.* **2002**, *354*, 518 and *J. Phys. Chem. A* **2001**, *105*, 10952; Tauber, M. J.; Mathies, R. A. *J. Am. Chem. Soc.* **2003**, *125*, 1394.

(52)   (a) Serra, S.; Iarlori, S.; Tosatti, E.; Scandolo, S.; Righi, M. C.; Santoro, G. E. *Chem. Phys. Lett.* **2002**, *360*, 487; (b) Sera, S.; Tossatti, E.; Iarlori, S.; Scandolo, S.; Santoro, G. *Phys. Rev. B* **2000**, *62*, 4389.

**Figure captions.**

**Fig. 1.**

Conceptual drawing of (a) the solvated electron in a polar solvent, (b) encapsulated electron in the cavity of a cage with polar groups (such as -*OH*) arranged inside, and (c) the same as (b), with polar groups (-*CN*) arranged outside. The arrow symbolizes the positive end of the electric dipole pointing to the electron.

**Fig. 2.**

(a) Decay kinetics $\kappa(t)$ for bi- 248 nm photon induced transient conductivity signals from $N_2$-saturated *n*-hexane solutions (at 18 °C) containing (from top to bottom) 0, 6.7, 19, 52, 144, 380, and 1050 µmol/dm$^3$ glutaronitrile. The solid lines are multi- trace least squares fit obtained using eq. (7S) in the Appendix. (b) Symbols: decay kinetics of 1064 nm induced "spike" in the photoconductivity ($\Delta\kappa$) obtained from the same solutions after laser excitation at $t_L \approx 62$ ns. The same concentrations as in (a), plotted from bottom to top. For convenience, the traces are stacked on top of each other. The solid lines drawn through the symbols are the least squares fits obtained using eq. (15S). (c) *Filled circles:* concentration plot for the reciprocal time $\tau_{eq}^{-1}$ of settling equilibrium reaction (2) as determined from the analysis of kinetics shown in (b). [*S*] is the concentration of glutaronitrile, in µmol/dm$^3$. The slope of the line drawn through the symbols gives the rate constant of forward reaction (2).

**Fig. 3.**

Decay kinetics of photoconductivity signal $\kappa(t)$ from $N_2$-saturated solution of succinonitrile following bi- 248 nm photon ionization of the solvent, at (a) 18 °C, (b) 26 °C, (c) 33.5 °C, and (d) 42 °C. The concentrations of the solute, in µmol/dm$^3$, were (from the top trace to the bottom trace in each series): (a) 0, 0.83, 1.67, 3.3, 6.7, 13.7, and 27.2,

39.

(b) 0, 2.8, 5.8, 12.4, and 27, (c) 0, 3, 5.9, 11.6, and 27, and (d) 0, 3.1, 6.2, 12.5, and 27. The solid lines are multi- trace fits obtained using eq. (7S) in the Appendix.

**Fig. 4.**

The same as Figure 3, for butyronitrile at (a) 18.5 °C, (b) 8 °C, and (c) -3 °C. The concentrations of the solute were (from the top trace to the bottom trace in each series): 0, 3, 6, 12, and 23 mmol/dm$^3$. The solid lines are exponential fits; extrapolation of these fits to $t = 0$ gives $\kappa_0$; the equilibrium constant $K_{eq}$ of reaction (2) can be obtained from the linear slope of the plot of $\kappa_0([S]=0)/\kappa_0$ vs. $[S]$, the molar concentration of butyronitrile.

**Fig. 5.**

van't Hoff plot for molar equilibrium constants $K_{eq}$ of reaction (2) vs. reciprocal absolute temperature $T$, in K, for butyronitrile (squares), acetonitrile (downturned triangles; data taken from ref. 15, section 1S) , adiponitrile (open circles), CH2N (upturned triangles), and succinonitrile (diamonds). The thermodynamic parameters for reaction (2) determined from these plots are given in Table 1.

**Fig. 6.**

Isodensity contour maps of SOMO for radical anions of (a) butyronitrile, (b) malononitrile, and (c) glutaronitrile; for optimized geometry in the BLYP/6-31+G** model (gas phase). The lobes of different sign are colored purple and pink, respectively. The isodensity contours correspond to (a) ±0.04, (b) ±0.06, and (c) ±0.04 e Å$^{-3}$. The geometry and magnetic parameters are given in Table 1S in the Supplement.

**Fig. 7.**

Same as Figure 6, for syn- and anti- conformations of the radical anion of succinonitrile (isodensity contours at ±0.05 e Å$^{-3}$ are shown).

**Fig. 8.**

Same as Figure 6, for (a) syn- and (b) anti- conformations of the radical anion of adiponitrile and (c) CH2N. The isodensity contours at ±0.04 e Å$^{-3}$ are shown.

**Fig. 9.**

Isodensity contour maps of SOMO for the pentamer radical anion of ethane; for optimized geometry in the BLYP/6-31+G** model (gas phase). Triangular dipyramid star shaped arrangement with $C_{3h}$ symmetry was assumed. The isodensity contours correspond to (a) ±0.01, (b) ±0.0175, (c) ±0.0185, and (d) ±0.03 e Å$^{-3}$. The geometry and magnetic parameters are given in Table 3. The cross in the middle is the geometric cavity center.

**Fig. 10.**

The same as Figure 9, for pentamer radical anion of acetonitrile. The isodensity contours correspond to (a) ±0.008, (b) ±0.015, (c) ±0.02, and (d) ±0.026 e Å$^{-3}$.

**Fig. 11.**

The same as Figure 9, for a radical anion that includes four molecules of ethane and one of acetonitrile (placed radially on the long axis). The isodensity contours correspond to (a) ±0.01, (b) ±0.0175, (c) ±0.025, and (d) ±0.04 e Å$^{-3}$.

**Fig. 12.**

The same as Figure 9 for radical anions of (a) 1,3,5,7-tetracyano-cyclooctane and (b) 1,2,4,5,7,8,10,11-octacyano-cyclododecane. The isodensity contours correspond to (a) ±0.025 and (b) ±0.02 e Å$^{-3}$. Both rings are nonplanar (the anions are shown from the top). See Figure 9S for a side view.

## Table 1

Experimental thermodynamic and kinetic parameters for reaction (2) in *n*-hexane.

| solute | $-\Delta G^0$, kJ/mol | $-\Delta S^0$, J/mol.K | $-\Delta H^0$, kJ/mol | $-\Delta H^0$, meV | $K_{eq}$,[a] $M^{-1}$ | $k_2$ [a] x $10^{11}$ $M^{-1} s^{-1}$ | $k_3$ [a] x $10^{10}$ $M^{-1} s^{-1}$ |
|---|---|---|---|---|---|---|---|
| acetonitrile | 15±2 | 15.4±3 | 20±1 | 200 | 470 | - | - |
| butyronitrile | 12.5±5 | 17±8 | 17±2 | 180 | 150 | - | - |
| adiponitrile | 24±4 | 61±6 | 42±2 | 435 | 16400 | 4.1 | 0.3 |
| *trans*-1,4-cyclohexane dinitrile | 23±7 | 142±12 | 66±3 | 680 | 13000 | 6.8 | |
| glutaronitrile | 26.6 | - | - | - | 46200 | 6.8 | 1.7 |
| succinonitrile | 35±4 | 101±7 | 66±2 | 680 | $1.3 \times 10^6$ | 22 | 2.0 |

a) at 295 K.

## Table 2.

Calculated structural parameters, electron affinity, and free energy of solvation of mono- and di- nitriles and their anions. [a]

| solute | r(C-N), Å [b] | ∠C-C-N, ° [b] | -EA, eV | -$\Delta G^0$, eV (anion) | -$\Delta G^0$, meV (neutral) | $EA_{liq}$, eV |
|---|---|---|---|---|---|---|
| acetonitrile | 1.205 | 143.4 *(142.4)* [c] | 1.00 | 1.07 | 62 | ≈0 |
| butyronitrile | 1.204 | 145.7 | 0.94 | 0.99 | 93 | ≈ -0.04 |
| adiponitrile | 1.192 | 154.5 *(157.2)* | 0.52 *(0.44)* | 0.88 | 173 | 0.188 |
| *trans*-1,4-cyclohexane dinitrile | 1.182 | 167.6 | 0.46 | 0.84 | 212 | 0.172 |
| glutaronitrile | 1.192 | 159.1 | 0.36 | 0.89 | 122 | 0.372 |
| succinonitrile | 1.205 *(1.195)* | 151.7 *(152.2)* | 0.34 *(0.19)* | 0.98 | 90 | 0.514 |
| malononitrile | 1.192 | 152.5 | 0.27 | 0.95 | 153 | 0.59 |

a) BLYP/6-31+G** calculation (ref. 41 and 42); solvation energies are estimated using IEFPCM method (ref. 46); all conformations for dinitriles are lower-energy *anti- (trans-)* unless in italics; b) in the same DFT model for neutral mono- and di-nitriles, the *length r(C-N)* of the *C-N* bond is 1.174 Å and the *C-C-N* angle is 180°; c) for monomer anion of acetonitrile with methyl hydrogen and cyanide nitrogen in *syn-* position.

## Table 3.

Geometry and magnetic parameters for three optimum geometry multimer radical anions (Figures 9, 10, 11).

|  | $(C_2H_6)_5^-$ | $(CH_3CN)(C_2H_6)_4^-$ | $(CH_3CN)_5^-$ |
|---|---|---|---|
| ∠C-C-N - 180° | - | 0.32 | 0.11 (0.27) |
| r(C-N) | - | 1.175 | 1.174 (1.175) |
| r(C-CN) | - | 1.472 | 1.471 (1.471) |
| r(C-H) | a 1.034 (1.1035) <br> b 1.044 (1.1044) | (ac.) 1.105 <br> a 1.033 (1.034) <br> b 1.041 (1.041) | 1.103 (1.104) |
| r(X-H) | a 4.7 (3.42) <br> b 7.05 (5.71) | (ac.) 4.45 [a] <br> 4.45 (3.5) | 4.22 (3.11) [b] |
| ∠C-C-H | a 111.6 (111.7) <br> b 111.4 (111.4) | 111.6 | 110.8 (111.05) |
| $A(^{14}N)$ | - | 2.25 | 0.56 (0.66) |
| $A(^{13}C_N)$ | - | 0.86 | 1.17 (1.39) |
| $A(^{13}CH_3CN)$ | - | 55.4 | 26.7 (35) |
| $A(^{13}C_a)$ | 19.6 (19.11) | 15.4 (17.9) | - |
| $A(^{13}C_b)$ | 5.6 (3.89) | 2.9 (1.42) | - |
| $A(^1H_3C)$ | a -0.69 (-0.7) <br> b -0.33 (-0.405) | a -0.53 (-0.58) <br> b -0.25 (-0.21) | -0.47 (-0.54) |
| $A(^1H_3C-CN)$ | - | -1.0 | -0.47 (-0.54) |
| $2T(^{14}N)$ | - | 0.11 | 0.07 (0.08) |
| $2T(^{13}CN)$ | - | 0.06 | 0.11 |
| $2T(^{13}CH_3CN)$ | - | 0.15 | 0.14 (0.19) |
| $2T(^{13}C_a)$ | 0.51 (0.47) | 0.38 (0.39) | - |
| $2T(^{13}C_b)$ | 0.49 (0.41) | 0.33 (0.31) | - |
| $2T(^1H_3C)$ | a 0.73 (0.81) <br> b 0.39 (0.28) | a 0.57 (0.84) <br> b 0.22 (0.18) | - |
| $2T(^1H_3C-CN)$ | - | 1.33 | 0.68 (1.0) |

Bond distances and angles are given in Å and °, respectively. Parameters for axial atoms are given first, for atoms in the plane second (in parenthesis). For a neutral, $C_{3v}$ symmetrical acetonitrile molecule r(H-C)=1.1012 Å, r(C-C)=1.4681 Å, r(C-N)=1.1736 Å and ∠C-C-H=110.3°. For a neutral, $D_{3d}$ symmetrical ethane molecule, r(H-C)=1.1029 Å, r(C-C)=1.5435 Å, and ∠C-C-H=111.37°. Symbols "a" and "b" relate to the innermost and outermost methyl groups in ethane molecules, symbol "ac" relates to the methyl group in acetonitrile molecule. a) r(X-C)=5.75 Å, r(X-N)=7.08 Å; b) r(X-C)=4.79 Å, r(X-N)=5.97 Å.

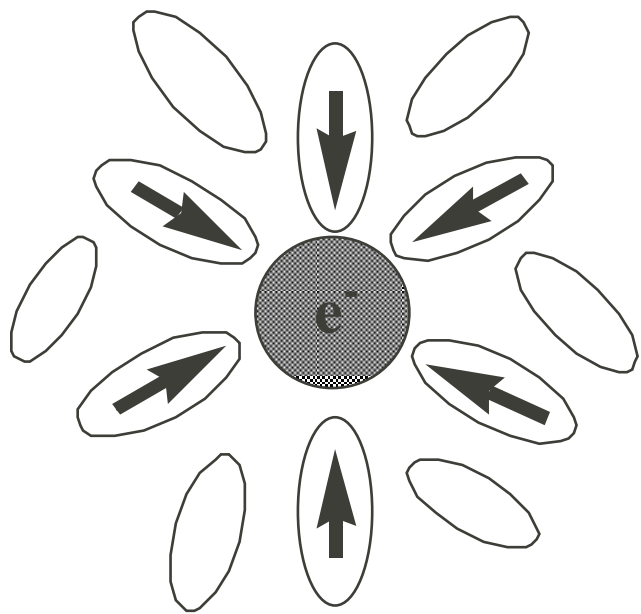
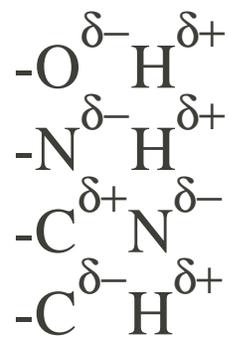
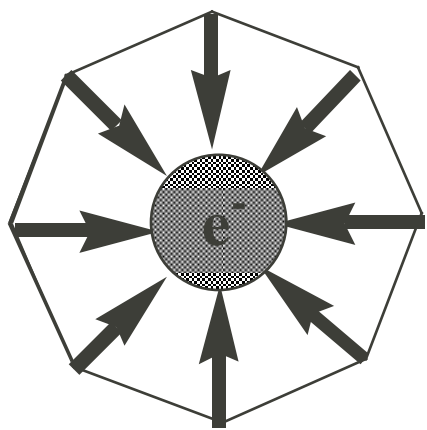
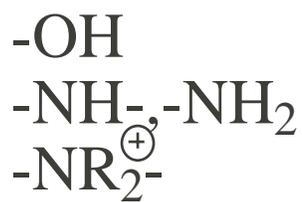
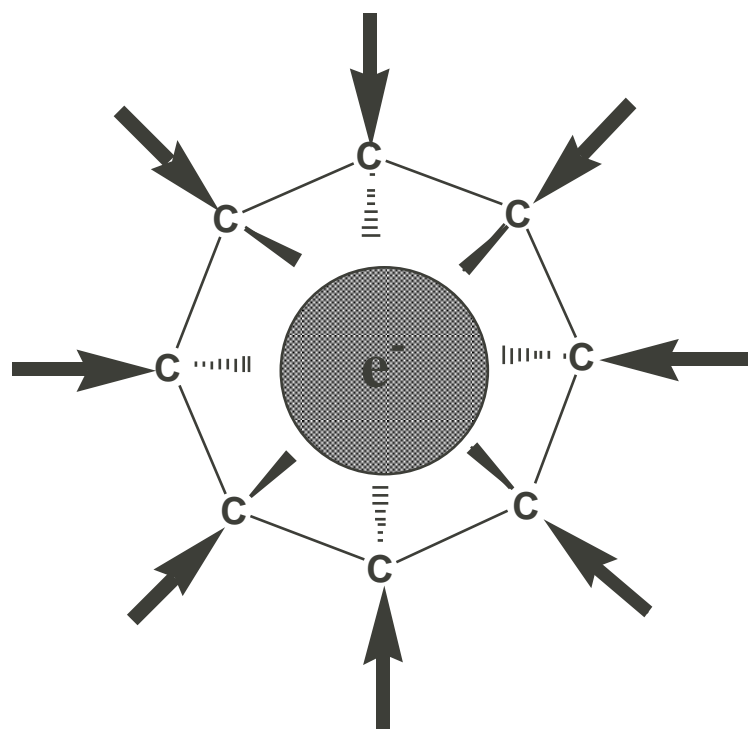
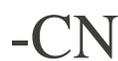

**Figure 1; Shkrob & Sauer**

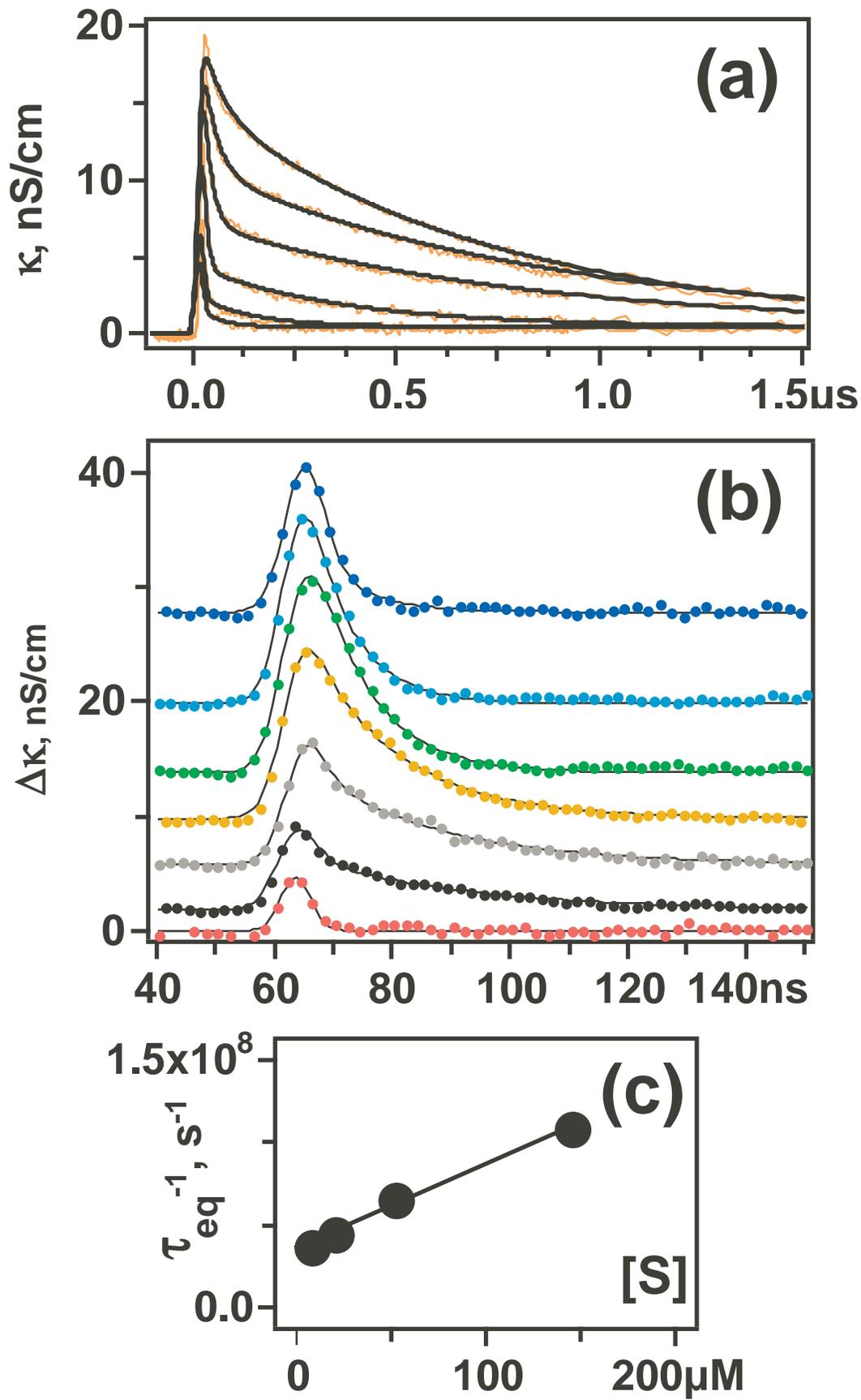

Figure 2; Shkrob & Sauer

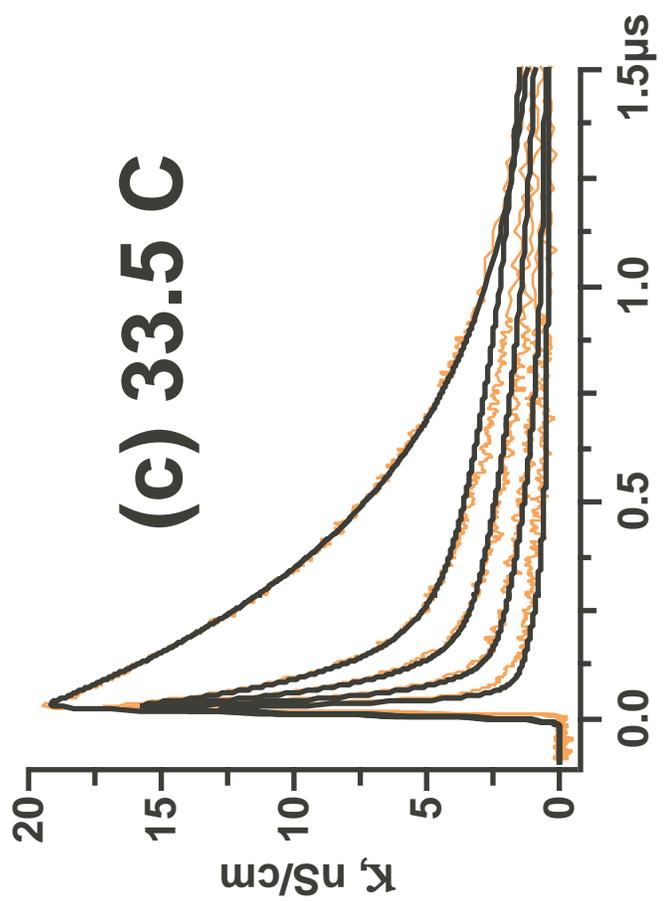
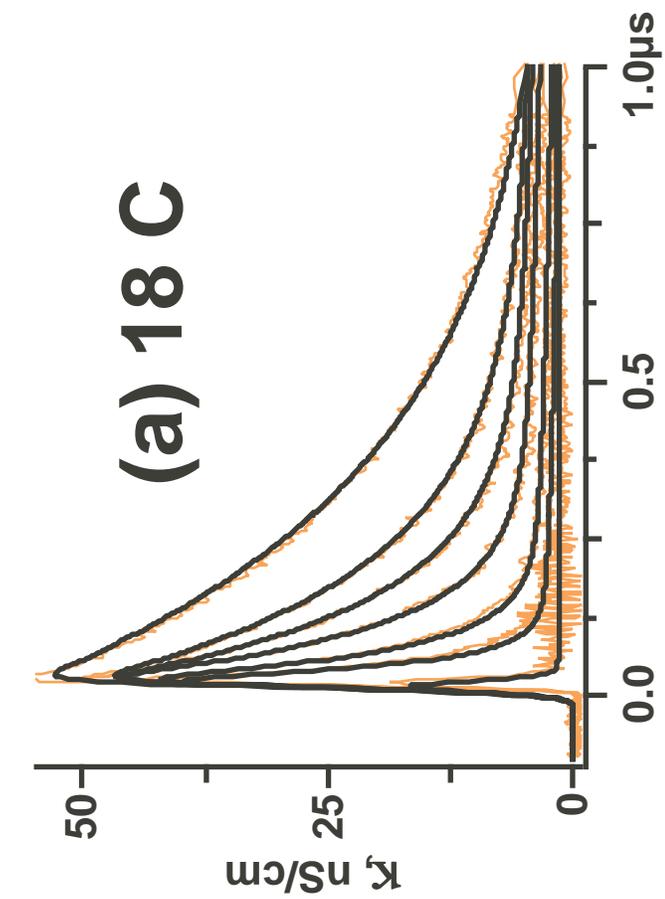
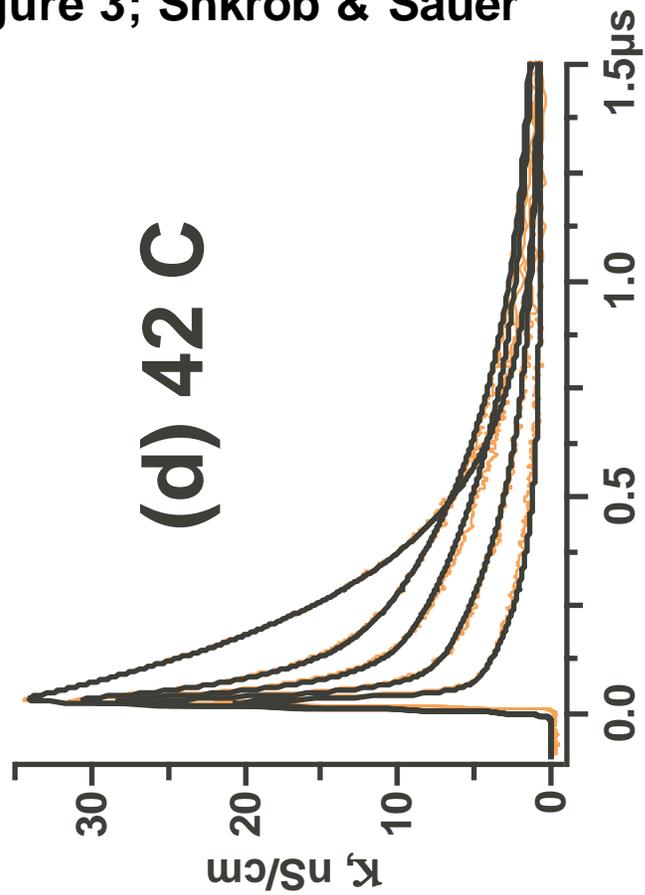
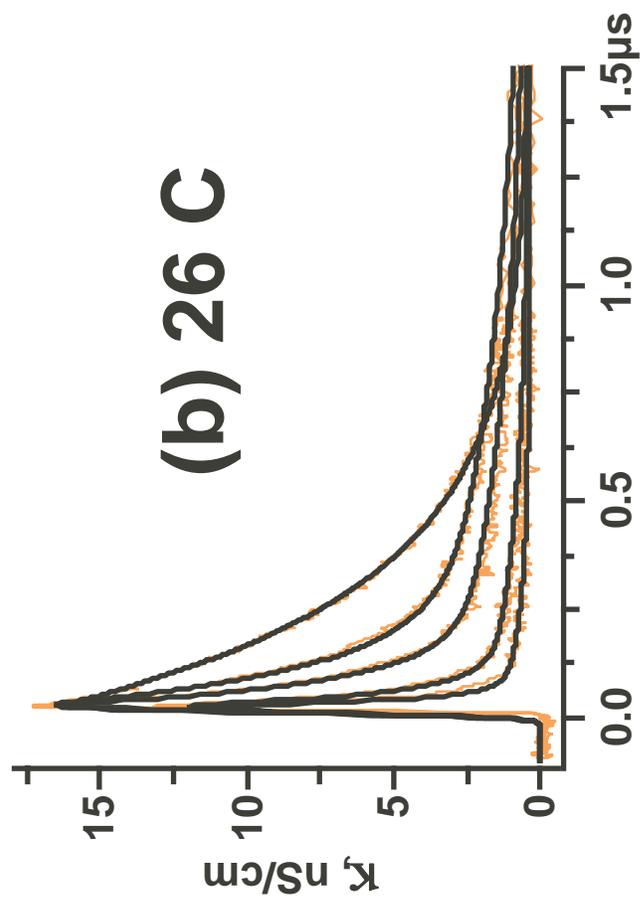

Figure 3; Shkrob & Sauer

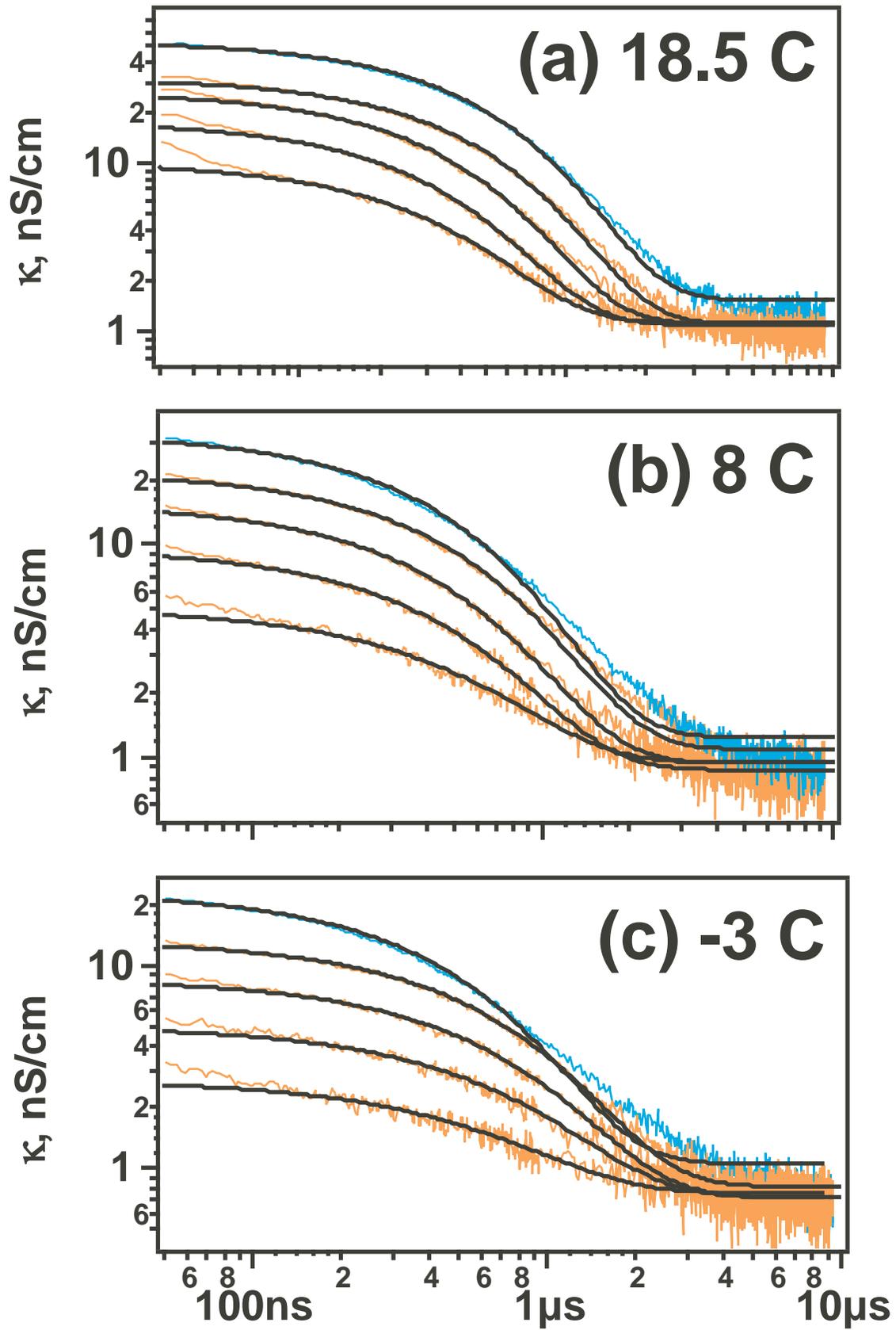

Figure 4; Shkrob & Sauer

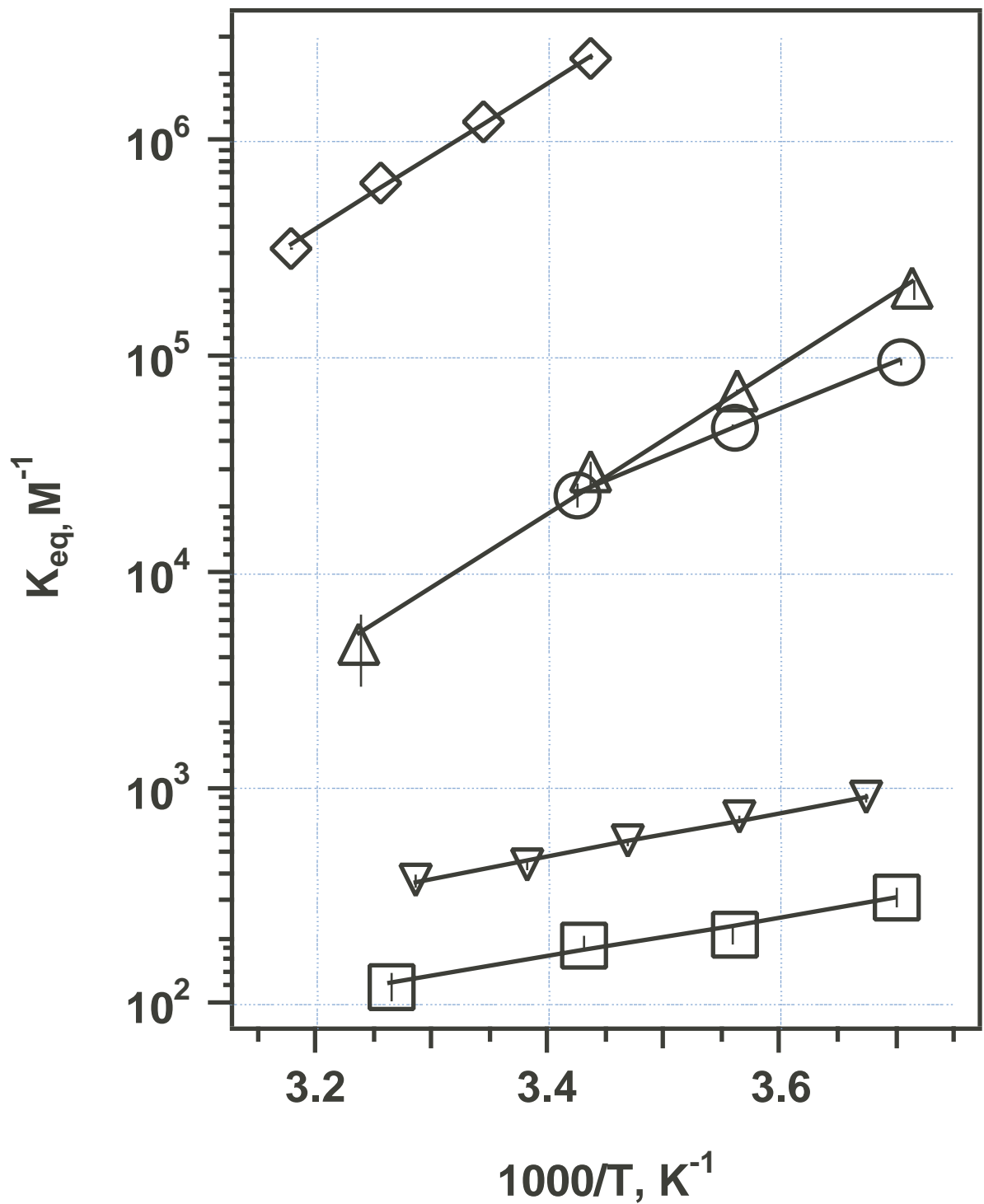

Figure 5; Shkrob & Sauer

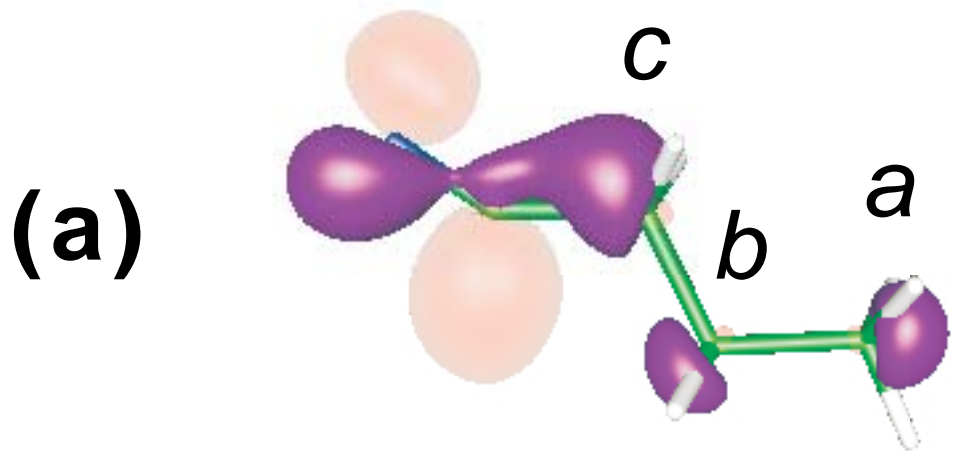

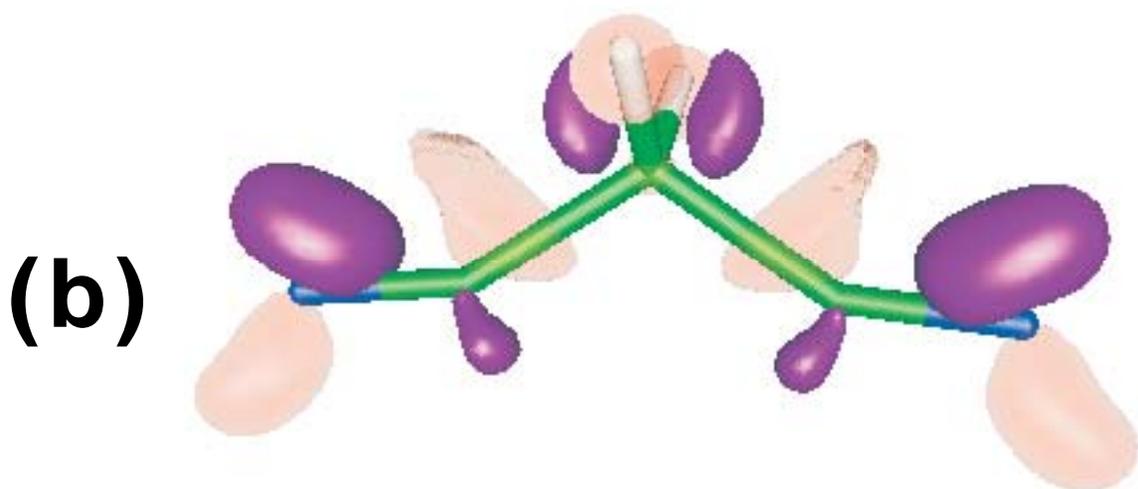

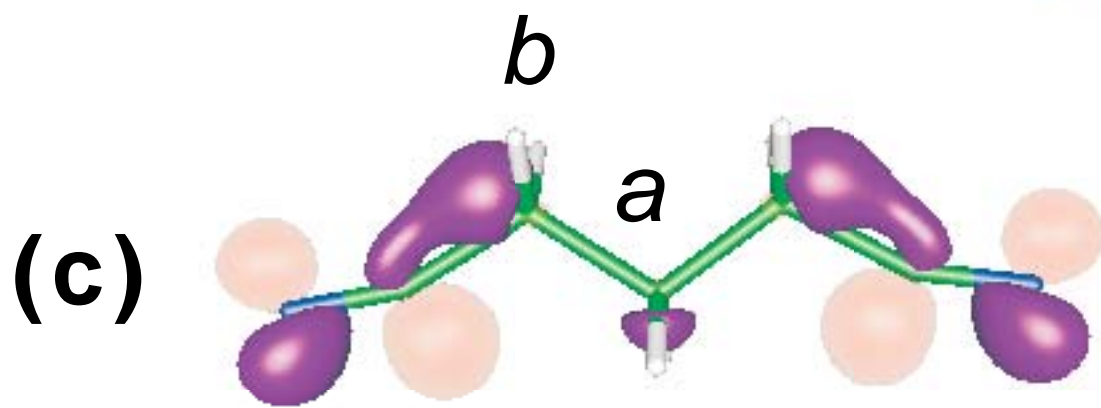

Figure 6; Shkrob & Sauer

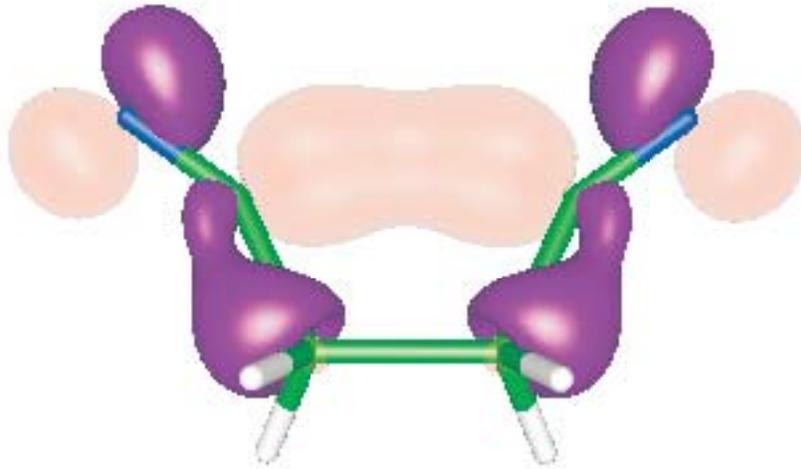

**(a)**

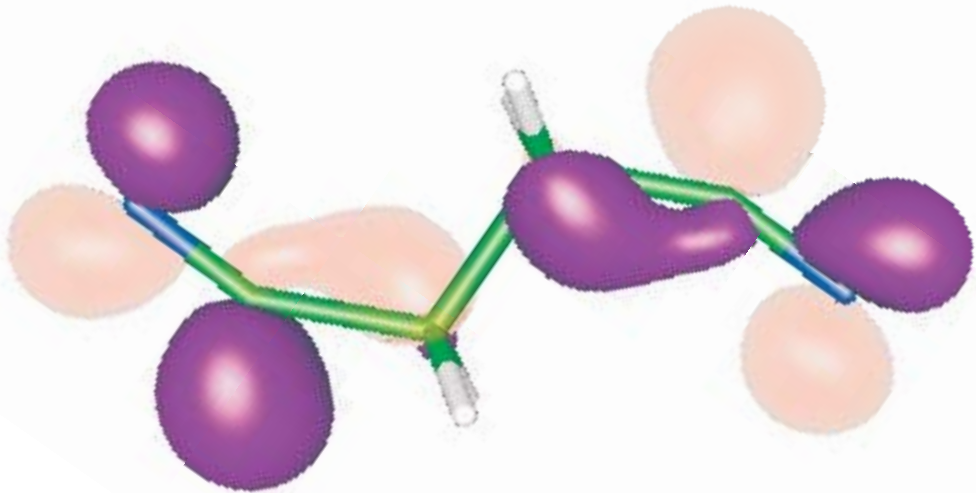

**(b)**

Figure 7; Shkrob & Sauer

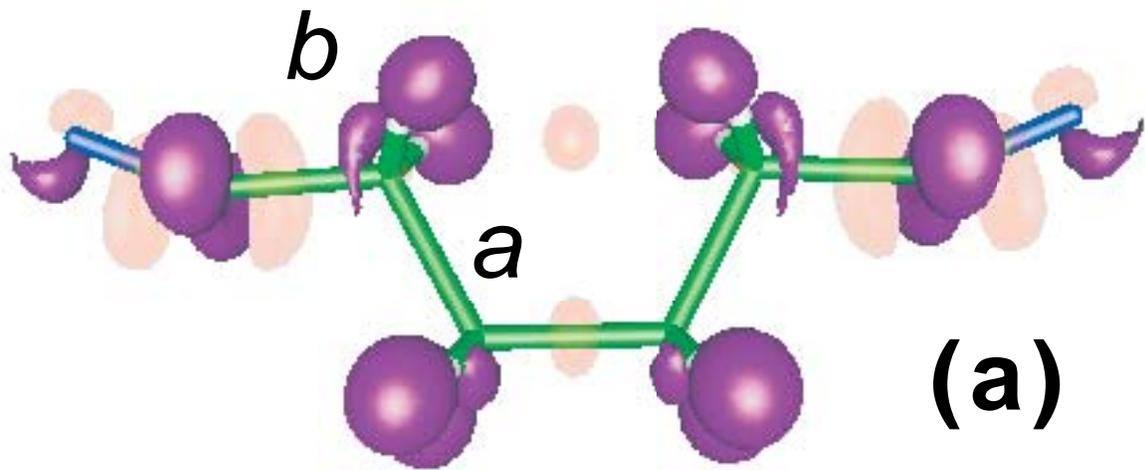
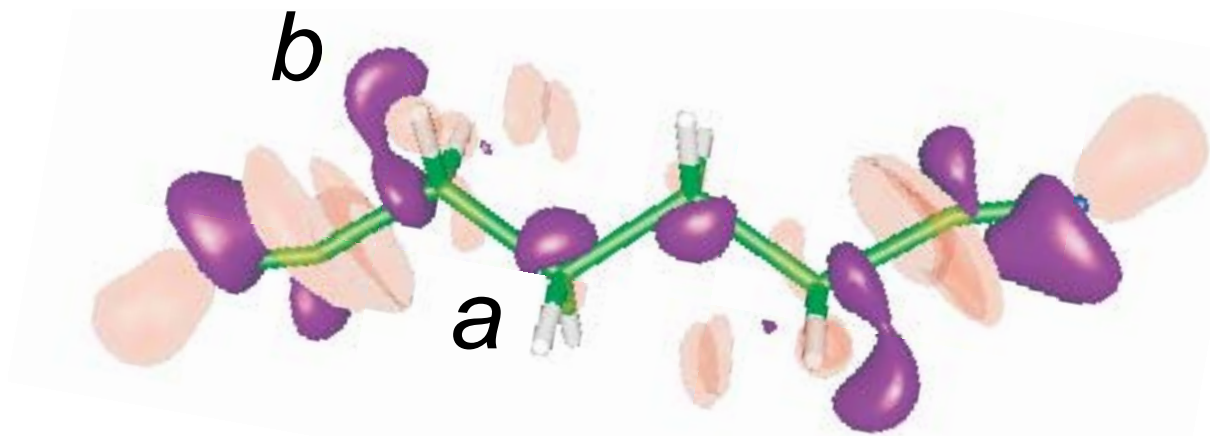
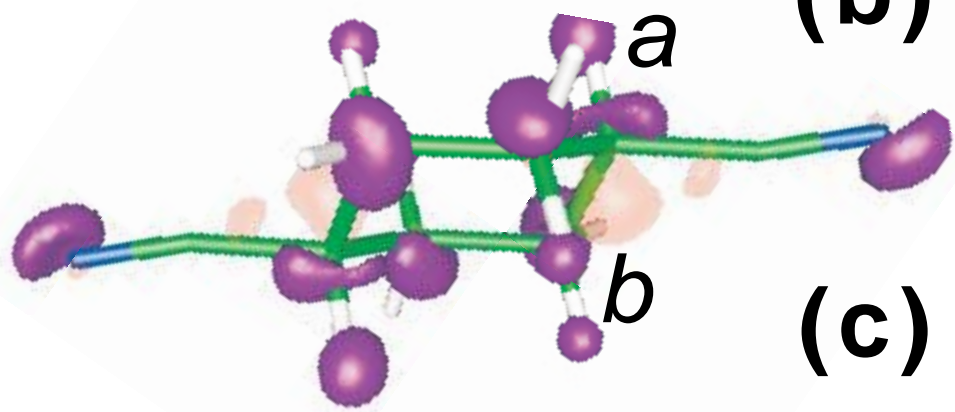

Figure 8; Shkrob & Sauer

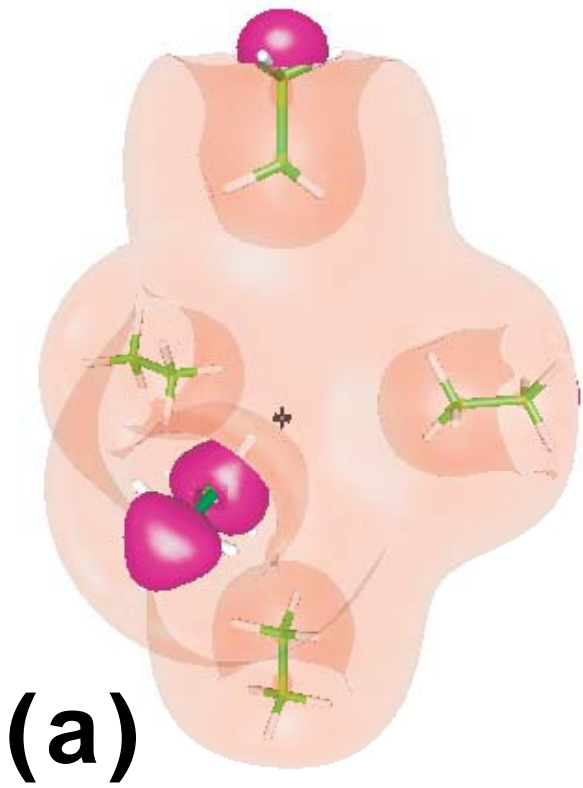 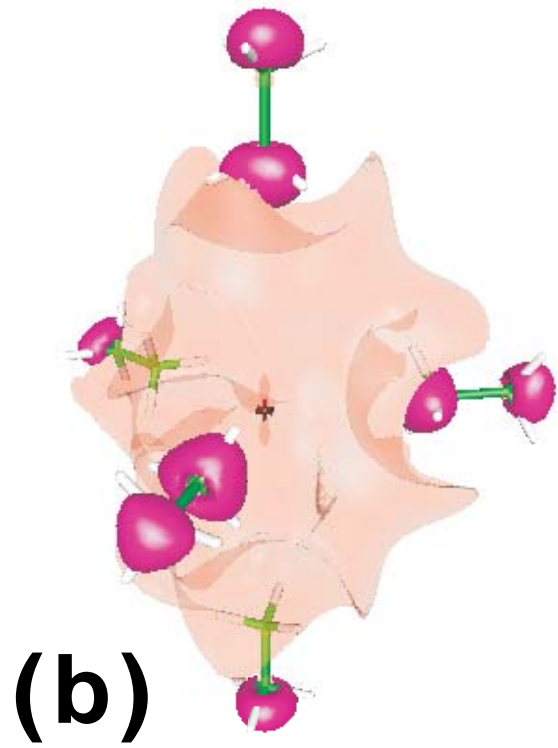
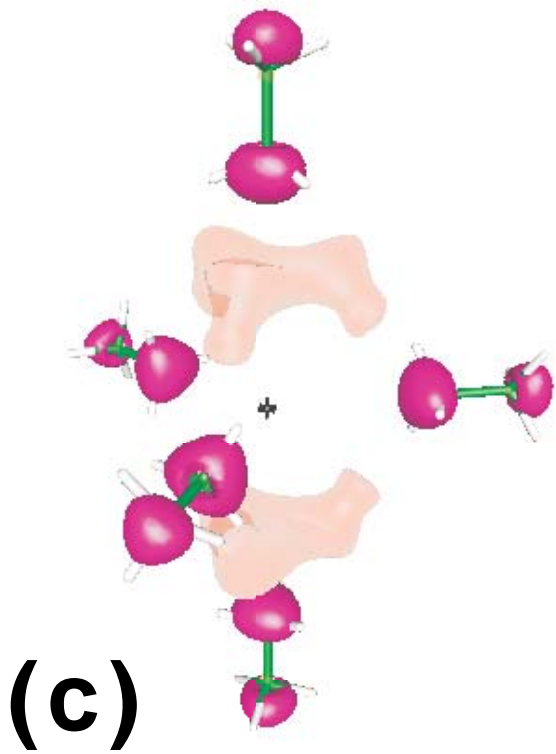 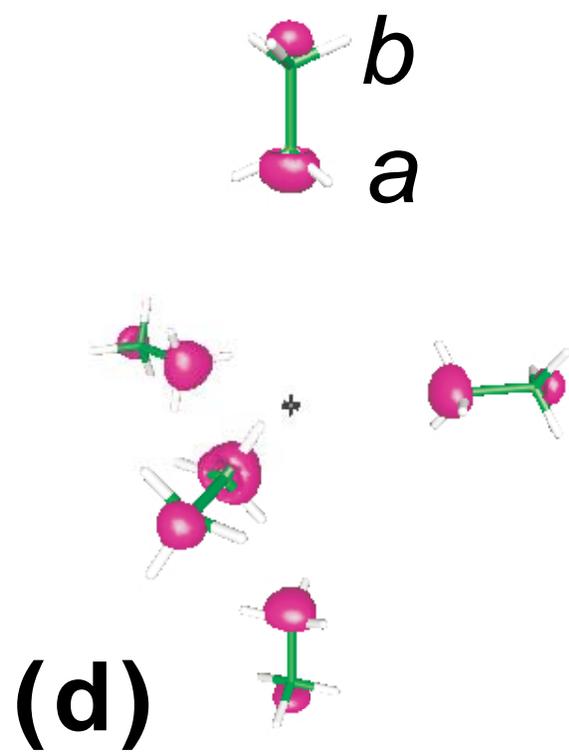

**Figure 9; Shkrob & Sauer**

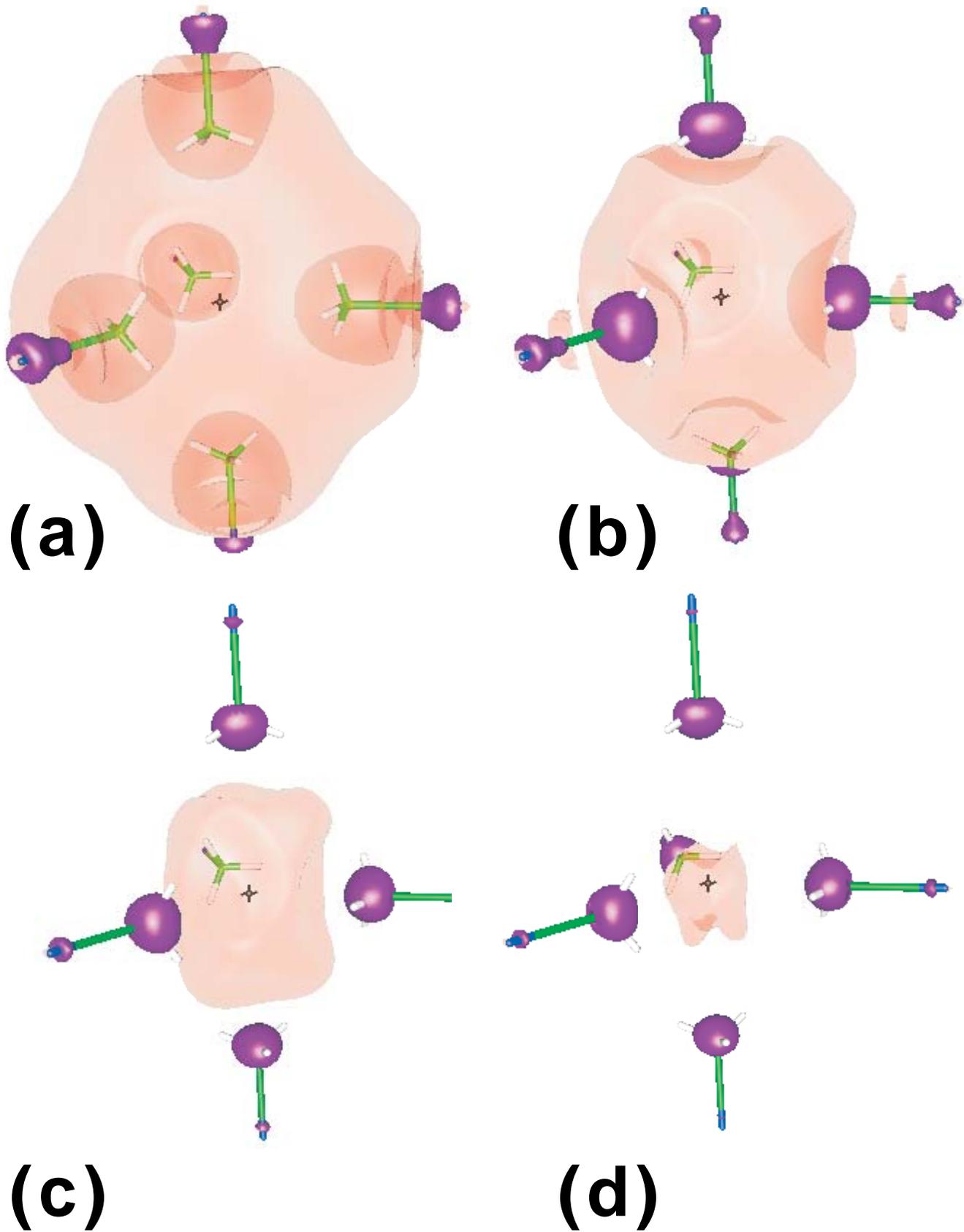

**Figure 10; Shkrob & Sauer**

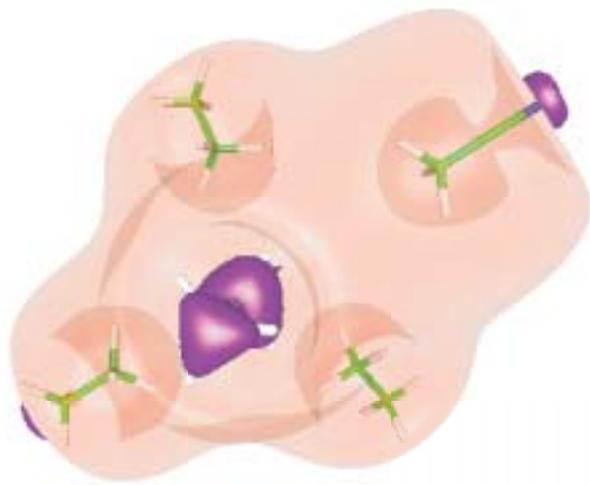
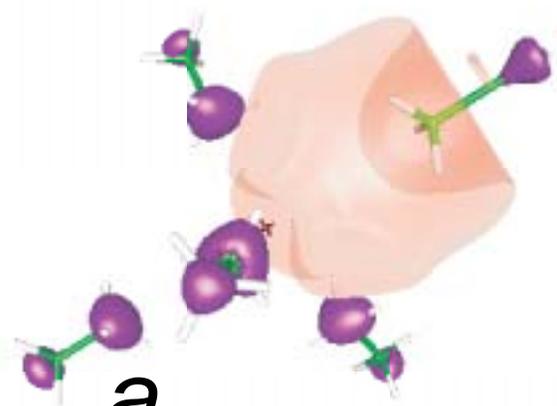

**(a)**      *b*   *a*     **(b)**

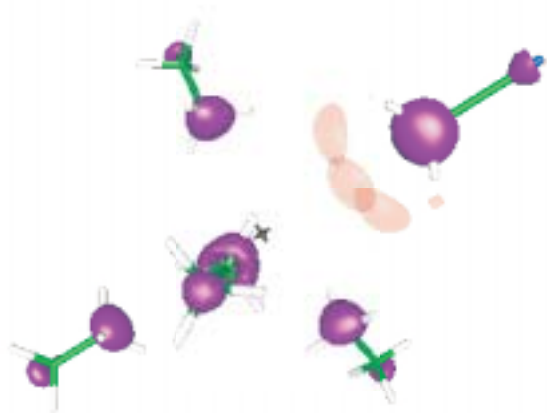
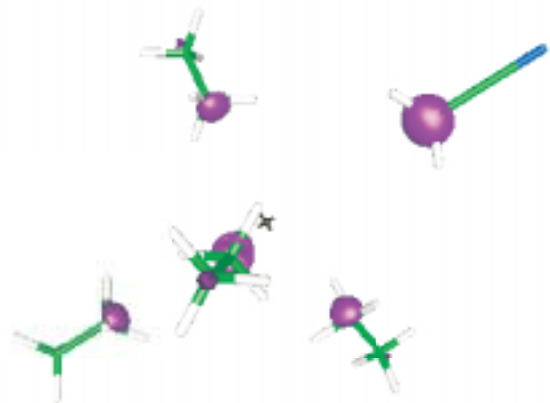

**(c)**        **(d)**

**Figure 11; Shkrob & Sauer**

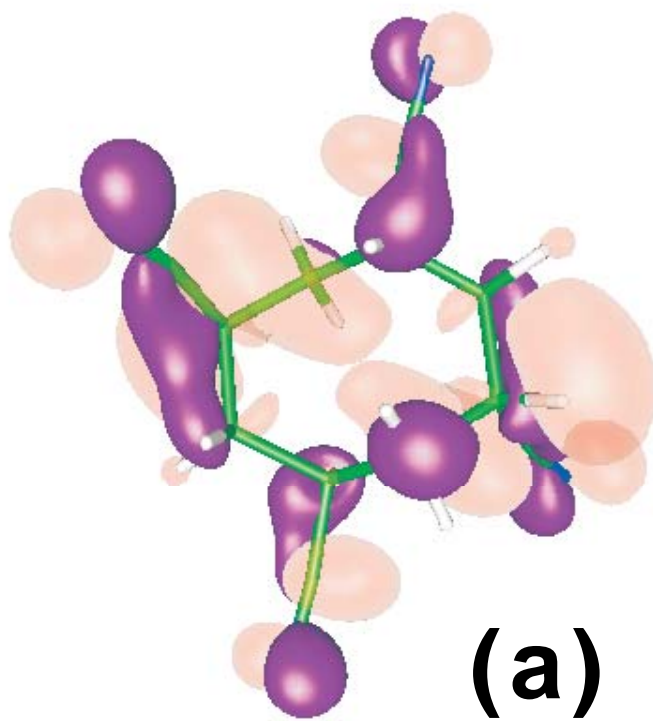

(a)

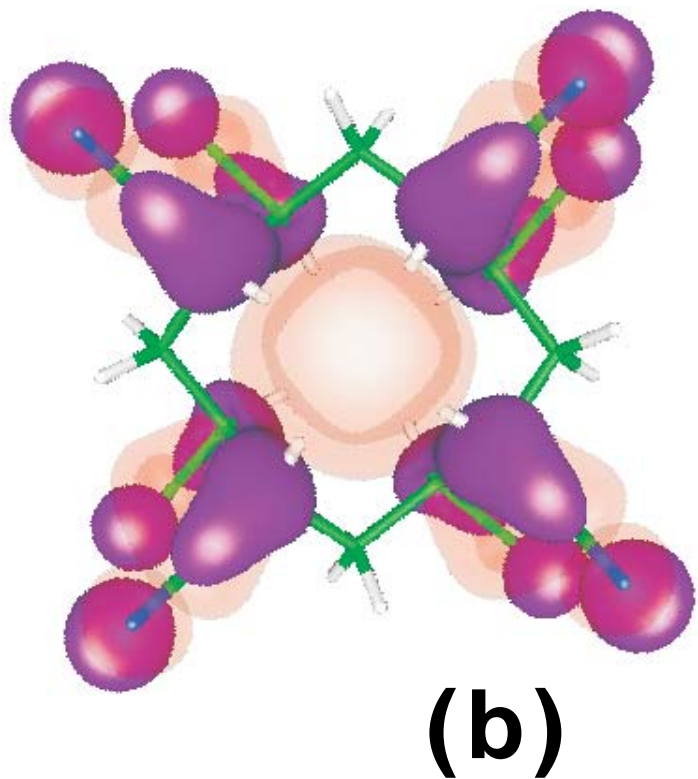

(b)

**Figure 12; Shkrob & Sauer**

**Towards electron encapsulation. Polynitrile approach.**

*Ilya A. Shkrob and Myran. C. Sauer, Jr.*

Radiation and Photochemistry Group, Chemistry Division, Argonne National Laboratory, 9700 South Cass Avenue, Argonne, Illinois 60439

*Tel* 630-2529516, *FAX* 630-2524993, *e-mail:* shkrob@anl.gov.

# Supporting Information.

**(1S.) Appendix. Details of kinetic analysis.**

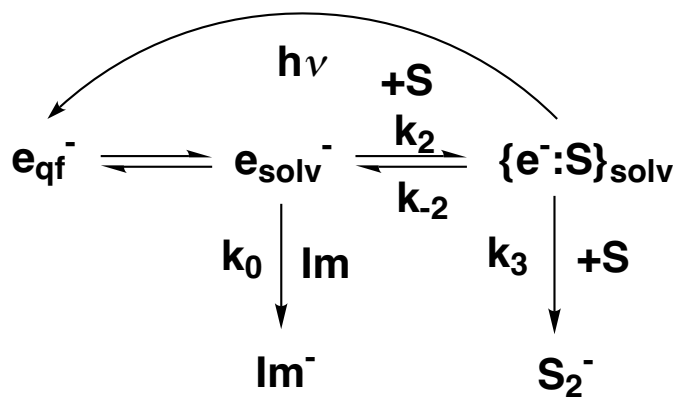

The analysis of electron equilibria kinetics followed the reaction scheme shown above. Introducing pseudo first order reaction constants

$$r_1 = k_2[S],\ r_2 = k_0,\ r_3 = k_{-2},\ r_4 = k_3[S], \tag{1S}$$

the observed rate constants $q_{1,2}$ of biexponential decay of $e^-_{solv}$ can be expressed through quantities

$$z = r_1 + r_2 + r_3 + r_4, \tag{2S}$$

$$v = r_2 r_3 + r_3 r_4 + r_4 r_1, \tag{3S}$$

$$d = \left[z^2 - 4v\right]^{1/2}, \tag{4S}$$

as

$$q_{1,2} = (z \mp d)/2 \tag{5S}$$

Introducing coefficients

$$c_1 = (r_2 + r_3 - q_1)/d \tag{6S}$$

and $c_2 = 1 - c_1$, the simulated kinetics is given by a biexponential curve convoluted with a Gaussian laser pulse

$$\kappa(t) = G(t) \otimes \left\{ \kappa_i + \kappa_0 \left[ c_1 \exp(-q_1 t) + c_2 \exp(-q_2 t) \right] \right\}, \tag{7S}$$

where

$$G(t) \propto \left( \pi \tau_p \right)^{-1} \exp(-t/\tau_p) \tag{8S}$$

and $\tau_p \approx 13$ ns is the width of the 248 nm pulse. Note that we neglected geminate contribution to the electron dynamics (due to the rapid diffusional escape of the quasifree electron from the Coulomb field of the parent hole). For every given temperature, the set of kinetics $\kappa(t)$ obtained for several concentrations $[S]$ of the solute were simultaneously fit using eq. (8S) by means of least squares optimization. This procedure typically resulted in a robust set of estimates for the rate constants. Typical examples of such multitrace fits are shown in Figures 2a, 3, 4S and 5S. Another way of estimating the equilibrium parameters is through the instantaneous unsettling of equilibrium reaction (2) by means of photoinduced electron injection into the conduction band after the 248 nm pulse by a short 1064 nm laser pulse:

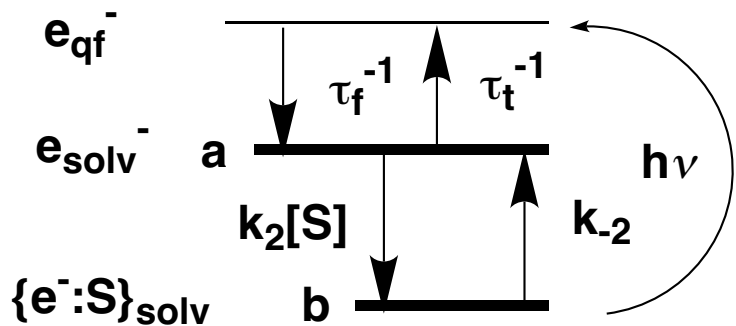

In the following, indexes "a" and "b" stand for the solvated electron $e^-_{solv}$ and $\{e^- : S\}_{solv}$ respectively. Let $\tau_f$ be the lifetime of the quasifree electron, $\tau_t$ the residence life-time of the solvated electron in a trap, $\sigma_{a,b}$ be the cross section for electron photodetachment from states "a" and "b", and $J(t)$ be the 1064 nm photon fluence. Assuming steady-state conditions for the quasifree electron, one obtains

$$\tau_f^{-1} \left[ e^-_{qf} \right] = \tau_t^{-1} a + \sigma_a J(t) a + \sigma_b J(t) b, \tag{9S}$$

2S.

and

$$db/dt = \sigma_b J(t)b + k_2[S]a - k_{-2}b. \tag{10S}$$

(Since $\left[e^-_{qf}\right] << a, b$, $da/dt \approx -db/dt$). Let $a_0$ and $b_0$ be the equilibrium concentrations of species "a" and "b" before the 1064 nm photon excitation, so that

$$a(t) = a_0 + \delta(t), \ b(t) = b_0 - \delta(t) \tag{11S}$$

and

$$d\delta/dt = \sigma_b J(t)b + \delta/\tau_{eq}, \tag{12S}$$

where

$$\tau_{eq}^{-1} = k_2[S] + k_{-2} \tag{13S}$$

is the reciprocal settling time of equilibrium reaction (2). For sufficiently weak laser pulses $\delta(t) << a_0, b_0$, we may assume $b \approx b_0$ and, therefore,

$$\delta(t) \approx \sigma_b b_0 J(t) \otimes \exp(-t/\tau_{eq}), \tag{14S}$$

so that the photoinduced change in the conductivity signal $\Delta\kappa(t) \propto \left[e^-_{qf}\right]_{\delta(t)} - \left[e^-_{qf}\right]_{\delta=0}$ is given by

$$\Delta\kappa(t) \propto [\sigma_a a_0 + \sigma_b b_0] J(t) + \tau_t^{-1} \sigma_b b_0 \ J(t) \otimes \exp(-t/\tau_{eq}). \tag{15S}$$

The decay kinetics of this 1064 nm photon induced change in the conductivity is, therefore, a weighted sum of the laser pulse $J(t)$ and this pulse convoluted with an exponential (for which the time constant equals the settling time of equilibrium reaction (2)). This time constant can be found by least squares fit of $\Delta\kappa(t)$ by eq. (15S) (as done, for example, in Figure 4S); the correlation of $\tau_{eq}^{-1}$ vs. $[S]$ (like that shown in Figure 2c) yields reaction constant $k_2$. In Figures 2b and 2c, this method is shown for glutaronitrile in room-temperature *n*-hexane. Formula (15S) is correct only when the equilibrium reaction (2) settles on the time scale that is much shorter than the "natural" life time of species "a" and "b" (determined by their reactions with the impurity and the solute, respectively). If that is not the case, a more complex equation has to be used: $\Delta\kappa$ is a weighted sum of three terms that are proportional to $J(t)$ and $J(t)$ convoluted with $\exp(-q_{1,2}t)$, where $q_{1,2}$ are given by eq. (5S).

**Table 1S.**

Calculated hyperfine constants for $^1$H, $^{13}$C, and $^{14}$N nuclei in mono- and di- nitrile anions (BLYP/6-31+G**).

| solute | $^{14}$N | $^{13}$C$_N$ | $^{13}$C$_H$ | $^1$H |
|---|---|---|---|---|
| *Syn*-acetonitrile $C_s$ 12 meV | 6.3 (5.2) | 32.2 (6.7) | 125 (1.2) | 0.6 (3.3) <br> 2x −0.2 (3.2) |
| *Anti*-acetonitrile $C_s$ | 6.3 (6.0) | 34.3 (7.5) | 125.5 (1.4) | 5.7 (3.2) <br> 2x −1.9 (3.3) |
| Butyronitrile $C_s$ ($^2$A') | 5.7 (5.7) | 29.1 (7.0) | a 62.6 (3.1) <br> b 15 (1.0) <br> c 23.7 (1.1) | 1a -0.33 (1.3) <br> 2a -0.56 (1.3) <br> 2b −0.7 (1.0) <br> 2c −1.7 (3.2) |
| malononitrile | 4.2 (2.1) | 4.9 (3.5) | 128 (1.1) | 6.6 (2.2) |
| *Trans*-adiponitrile $C_{2h}$ ($^2$A$_g$) | 3.4 (3.2) | 12.6 (4.8) | 2a 19.1 (1.2) <br> 2b 29.7 (2.1) | 4a -0.53 (1.4) <br> 4b -0.59 (2.0) |
| *Cis*-adiponitrile $C_{2v}$ ($^2$A$_1$) 182 | 4.2 (3.3) | 17.9 (4.9) | 2a 11.7 (0.83) <br> 2b 55.8 (1.5) | 4a -0.67 (0.66) <br> 4b -0.87 (2.4) |
| *Trans*-1,4-cyclohexane dinitrile $C_{2h}$ | 2.67 (1.1) | 7.05 (2.2) | 2a 9.8 (1.3) <br> 4b 16.0 (0.88) | 2a −0.77 (1.71) <br> 4ax 0.34 (1.5) <br> 4eq 0.43 (0.88) |
| Glutaronitrile $C_{2v}$ ($^2$A$_1$) | 3.53 (3.4) | 10.7 (4.5) | 2a 39.6 (0.16) <br> 4b 34.6 (2.4) | -0.75 (1.0) <br> -1.01 (2.6) |
| *Trans*-succinonitrile $C_{2h}$ ($^2$A$_1$) | 4.1 (6.9) | 23.6 (6.6) | 70.9 (2.9) | -0.93 (4.5) |
| *Cis*-succinonitrile $C_{2h}$ ($^2$B$_u$) 93 | 3.73 (4.4) | 13.5 (4.5) | 63.3 (1.4) | -0.57 (2.3) |

All constants are given in units of Gauss (1 G = 10$^{-4}$ T); isotropic hfcc's are given first, the largest principal value of the anisotropic tensor is given next, in parentheses. Symmetry and group representation are given next to the name of the molecule; for high-energy conformers, the energy relative to the lowest state is also given in meV.

**Figure captions (1S to 6S)**

**Fig. 1S.**

Schematic depiction of several classes of halide anion receptors that are based on quaternized amines: (a) katapinands (the filled circle shows the intercalated halide anion, $n>11$), (b) octaazacryptand, (c) azacrown, (d) Schmidtchen's cation ($n$=10-12; see Schmidtchen, F. P.; Müller, G. *J. Chem. Soc. Chem. Comm.* **1984**, 1115)

**Fig. 2S.**

Some specific implementations of the cage architecture shown in Figure 1c. (a,b) Cyano derivatized cycloalkanes, (b) azacrown, and (d) macrobicycle.

**Fig. 3S.**

Structural formulas of mono- and di- nitriles discussed in this paper.

**Fig. 4S.**

A series of 1064 nm photon induced $\Delta\kappa(t)$ kinetics for CH2N in N$_2$-saturated *n*-hexane (at 18 $^o$C) for $t_L \approx 65$ ns. The concentrations $[S]$ of CH2N in μmol/dm$^3$ are given in the legend. The solid lines drawn through the symbols are least squares fits obtained using eq. (15S).

**Fig. 5S.**

Decay kinetics of the conductivity signal obtained in bi- 248 nm photon ionization of N$_2$-saturated *n*-hexane solutions of CH2N at (a) -3.8 $^o$C, (b) 7.7 $^o$C, (c) 18 $^o$C, and (d) 36 $^o$C. The concentrations of the solute, in μmol/dm$^3$, were (from the top to the bottom trace in each series): (a) 0, 8.7, 17.4, 41, and 84, (b) 20.5, 40.7, 82.3, and 168, (c) 0, 20, 40, 80, 160, 320, and 580, and (d) 0, 19.6, 41, 81, 164, and 328. The solid lines are multitrace fits obtained using eq. (7S).

**Fig. 6S.**

The same as Figure 5S, for adiponitrile at (a) 19 $^o$C, (b) 8 $^o$C, and (c) -3 $^o$C. The concentrations of the solute were (a) 0, 25, 107, 220, and 440, (b) 0, 14.4, 30, 60, and 174, and (c) 0, 9.7, 19, 44, and 86 μmol/dm$^3$.

**Fig. 7S.**

Isodensity contour maps of SOMO for the radical anion of 1,3,5,7-tetracyano-cyclooctane for optimized geometry in the BLYP/6-31+G** model (gas phase). The lobes of different sign are colored purple and pink, respectively. The isodensity contours correspond to (a) ±0.02, (b) ±0.04, and (c) ±0.06 e Å$^{-3}$. The magnetic parameters are given in Fig. 10S.

**Fig. 8S.**

The same as Figure 7S, for the radical anion of 1,2,4,5,7,8,10,11-octacyano-cyclododecane (top view).

**Fig. 9S.**

The same as Figure 8S, side view. Note the dumbbell shaped electron orbital filling the cavity.

**Fig. 10S.**

Isotropic hyperfine coupling constants (in Gauss) for the radical anions of (a) 1,3,5,7-tetracyano-cyclooctane ($C_i$ symmetry, $A_g$ representation) and (b) 1,2,4,5,7,8,10,11-octacyano-cyclododecane ($C_2$ symmetry, $A$ representation); optimized structures in BLYP/6-31+G** model. Black is for $^{13}$C, blue is for $^{14}$N, red is for $^1$H. Compare to Figure 10 in the text.

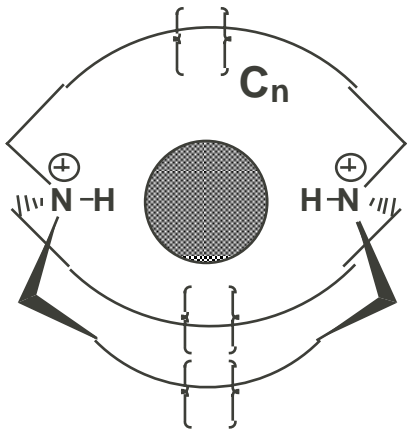
**(a)**

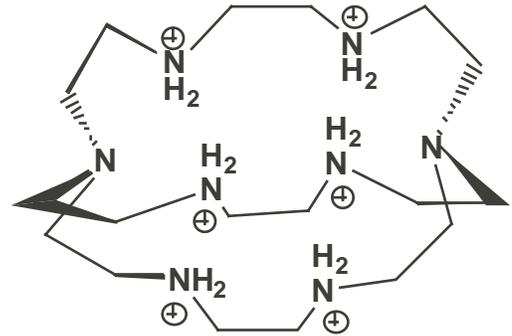
**(b)**

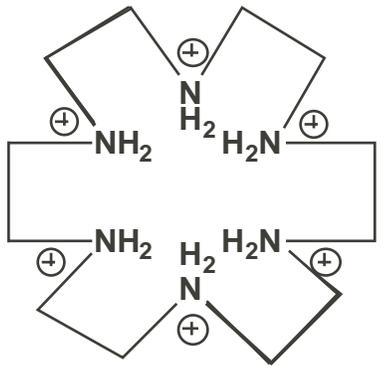
**(c)**

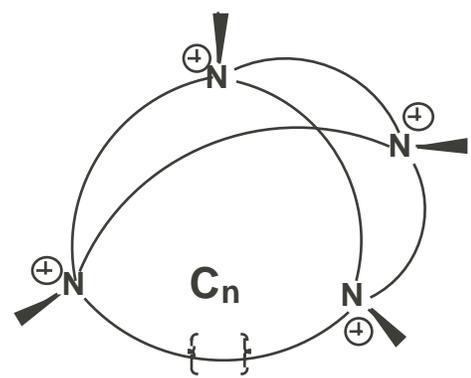
**(d)**

**Figure 1S; Shkrob & Sauer**

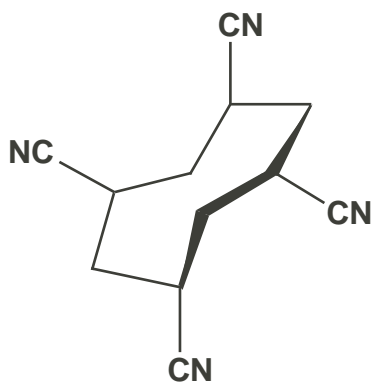 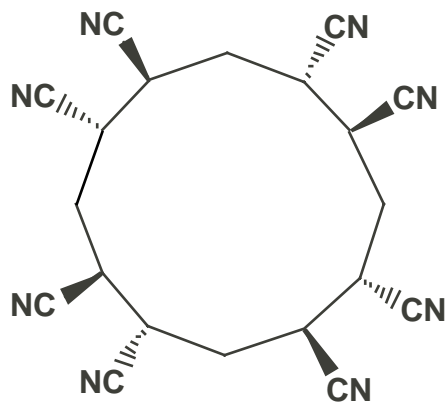

(a) (b)

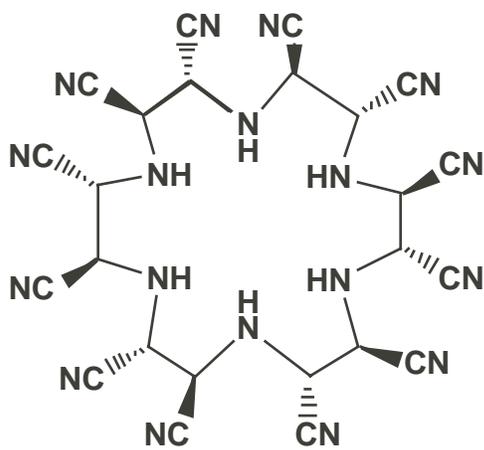 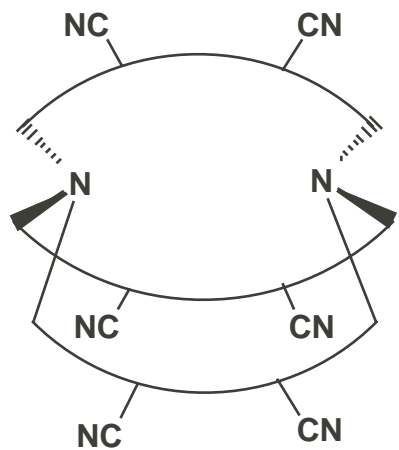

(c) (d)

**Figure 2S; Shkrob & Sauer**

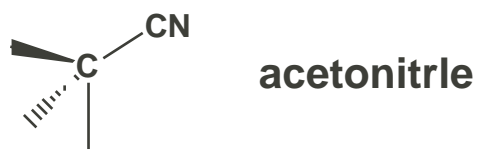 acetonitrle

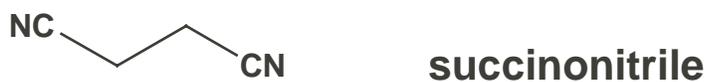 succinonitrile

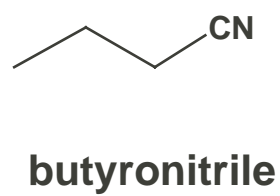
butyronitrile

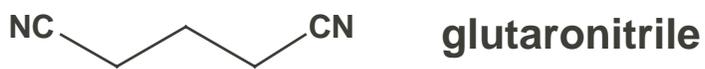 glutaronitrile

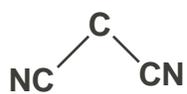
malonitrile

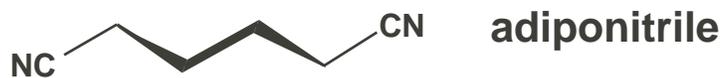 adiponitrile

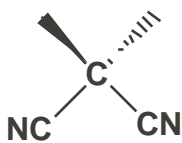
dimethyl-
malonitrile

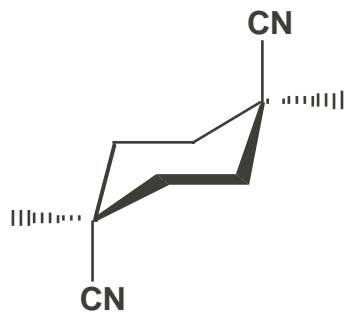
*trans*-1,4-cyclohexane
dicarbonitrile

**Figure 3S; Shkrob & Sauer**

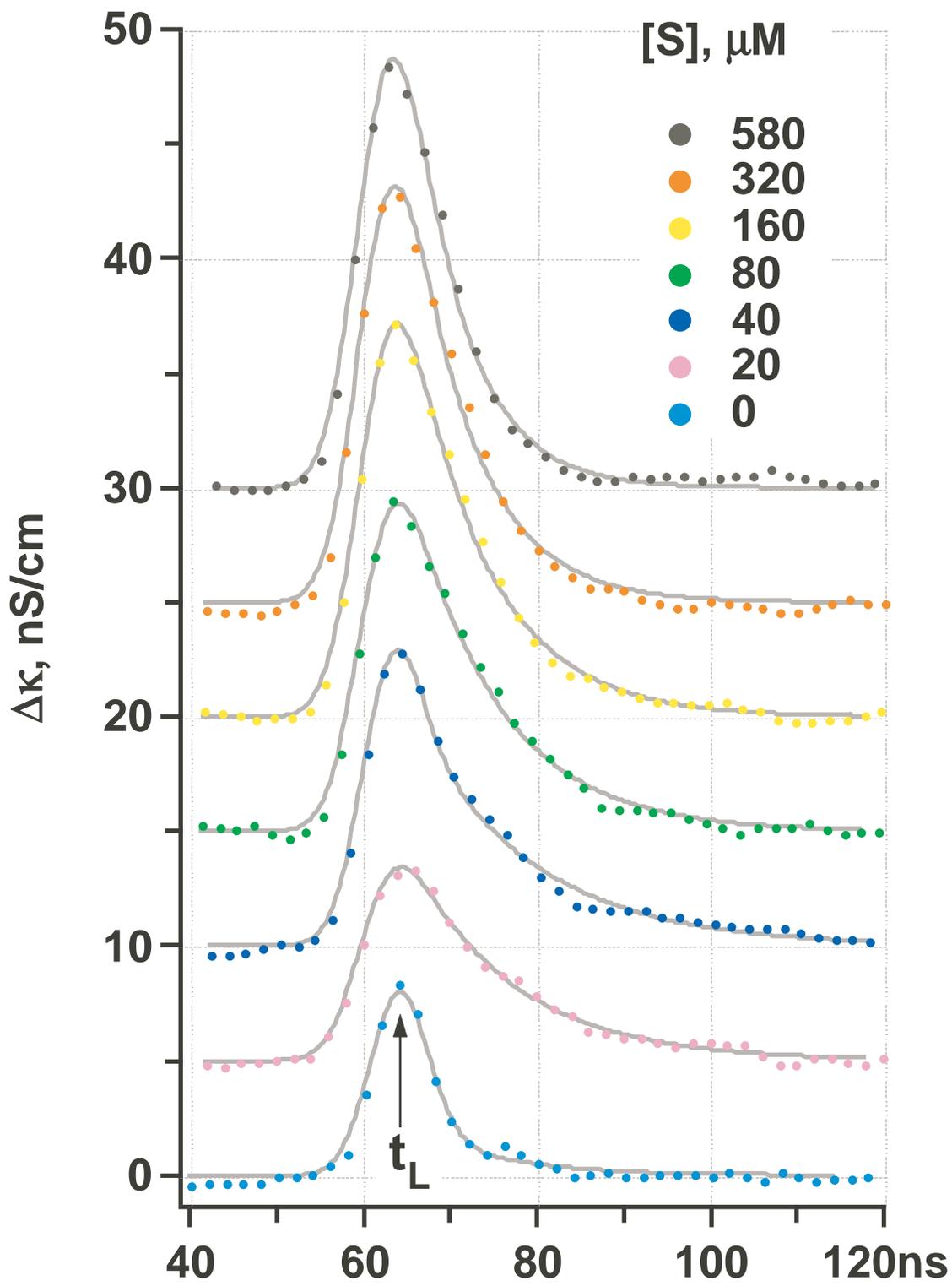

Figure 4S; Shkrob & Sauer

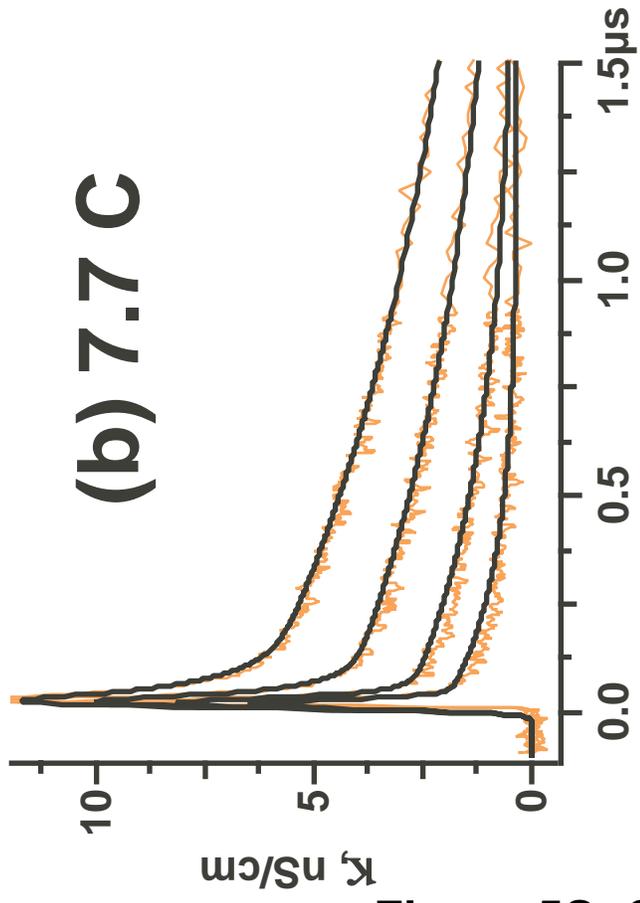
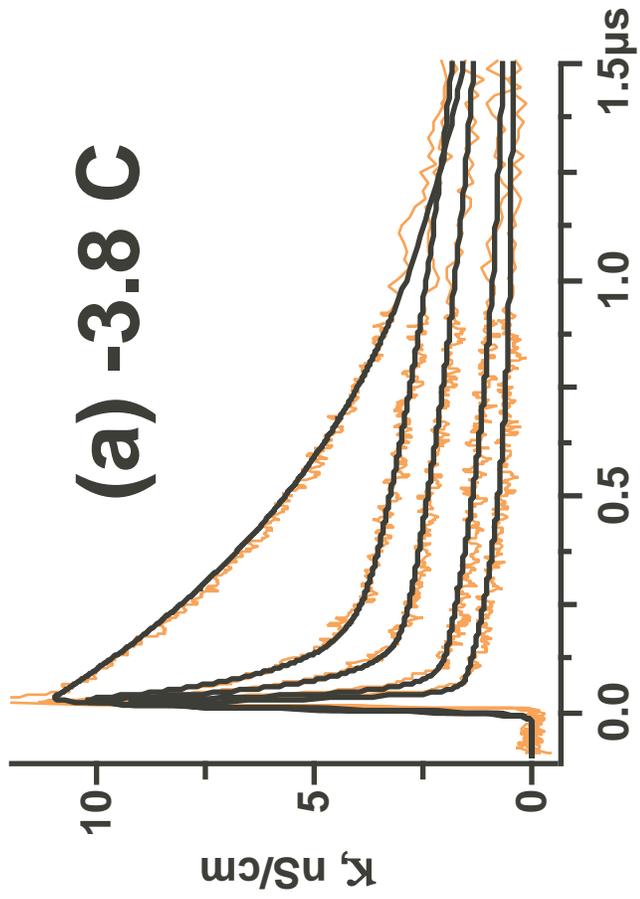
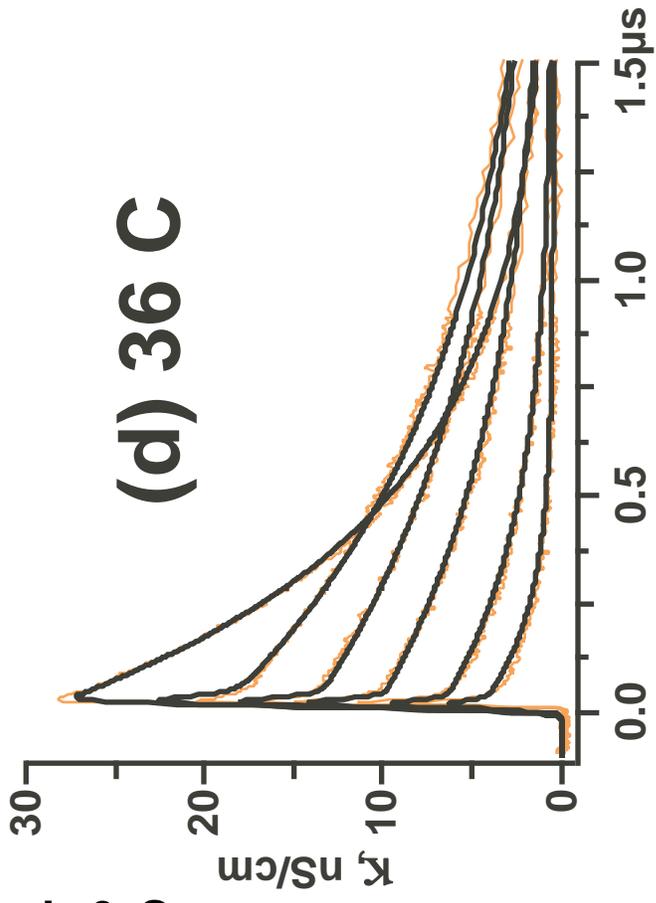
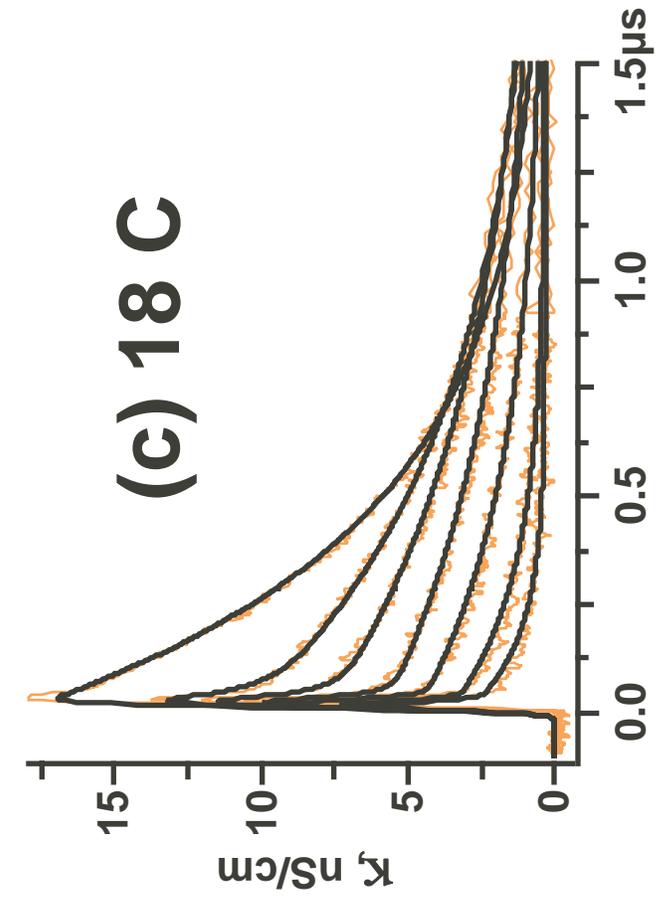

**Figure 5S; Shkrob & Sauer**

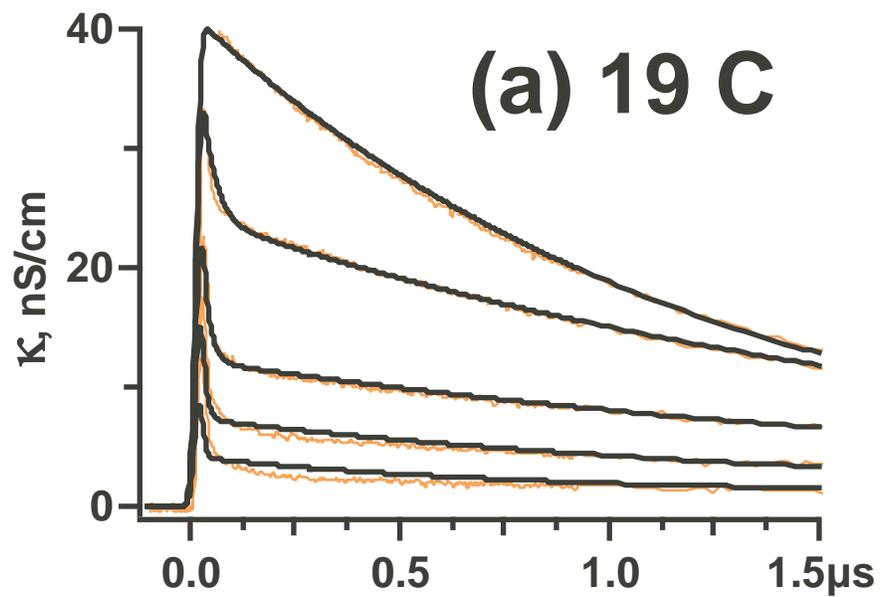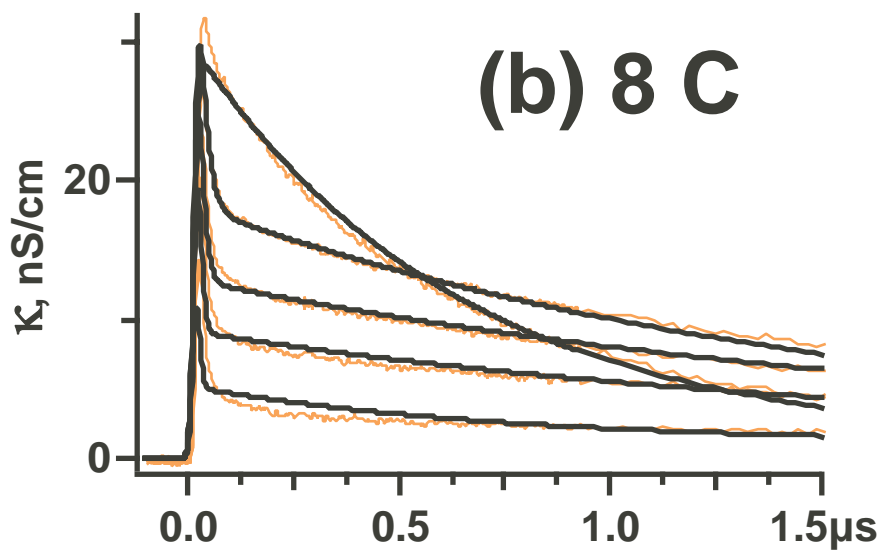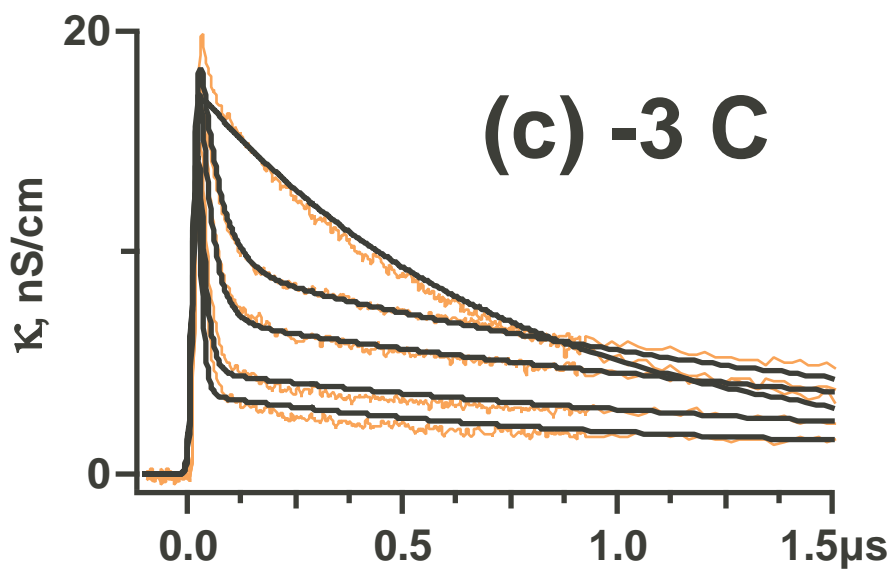

Figure 6S; Shkrob & Sauer

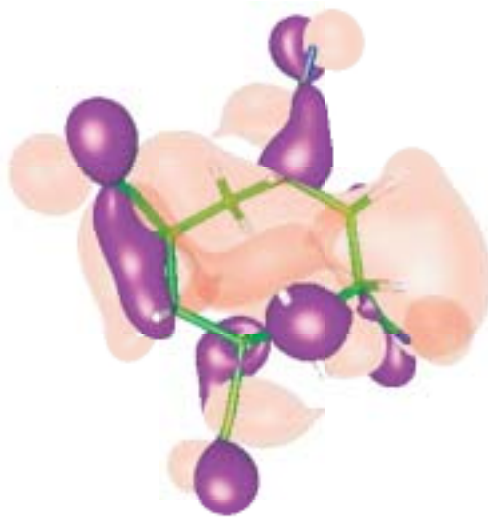

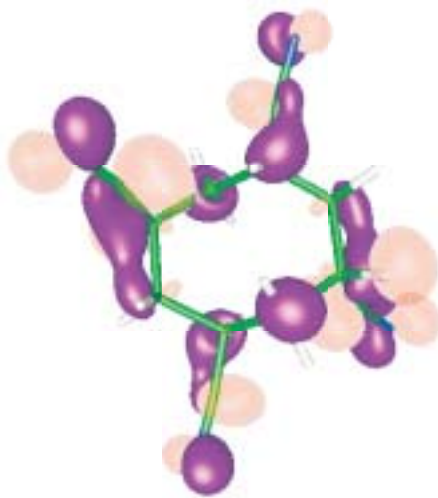

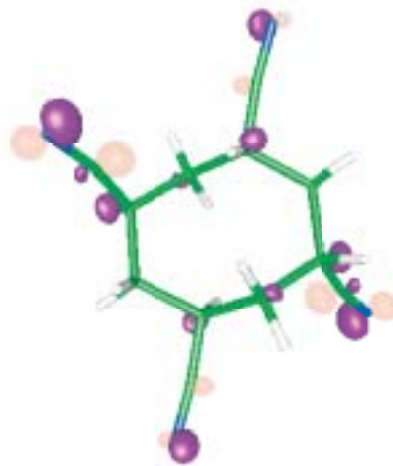

Figure 7S; Shkrob & Sauer

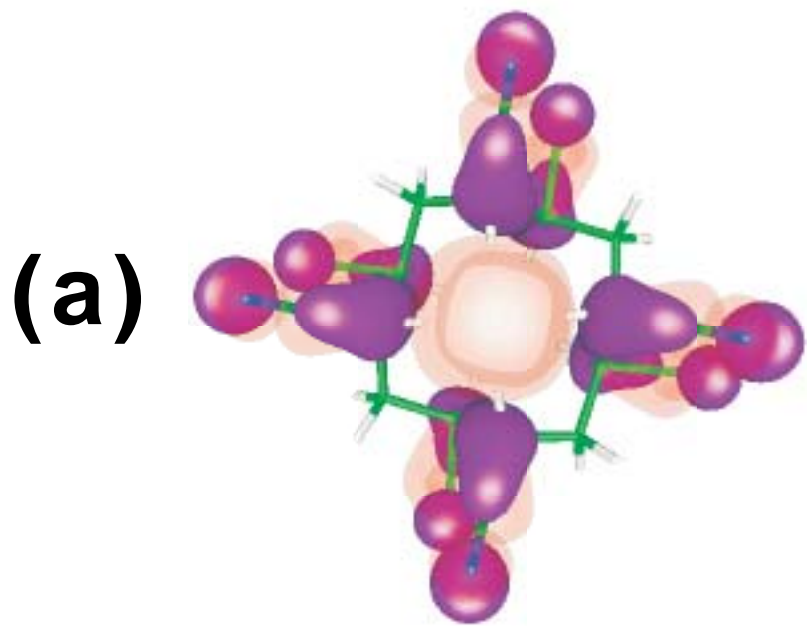

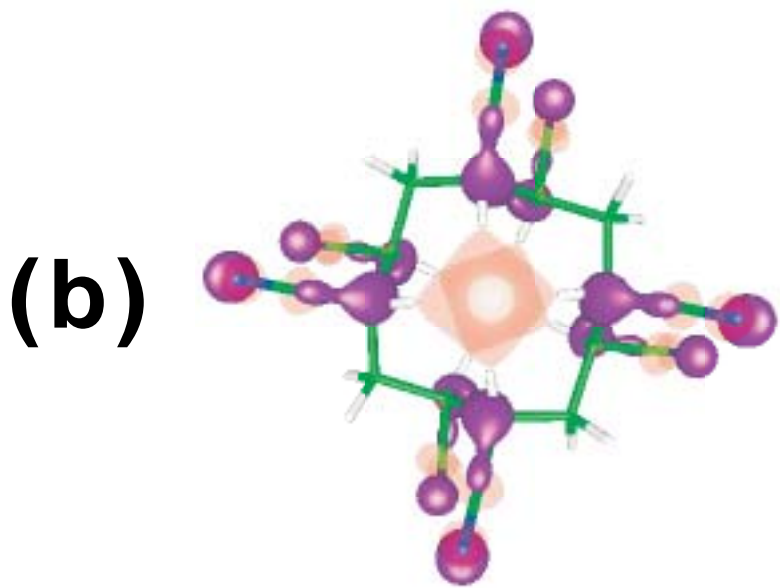

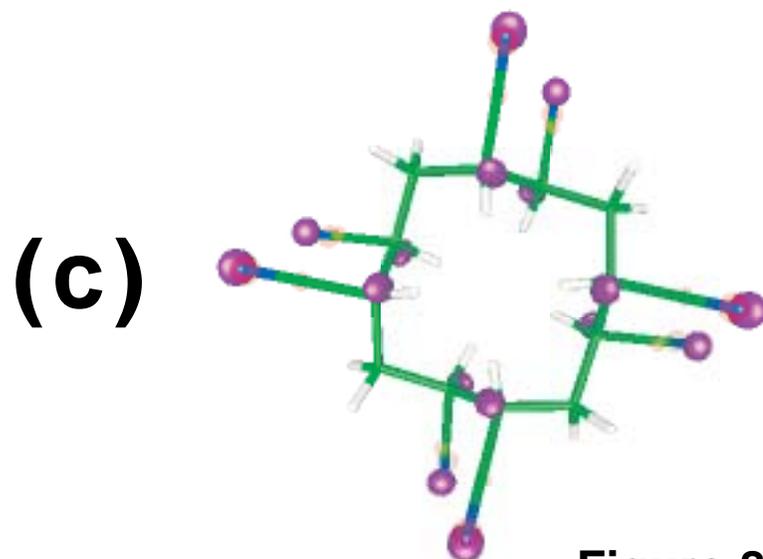

**Figure 8S; Shkrob & Sauer**

(a)

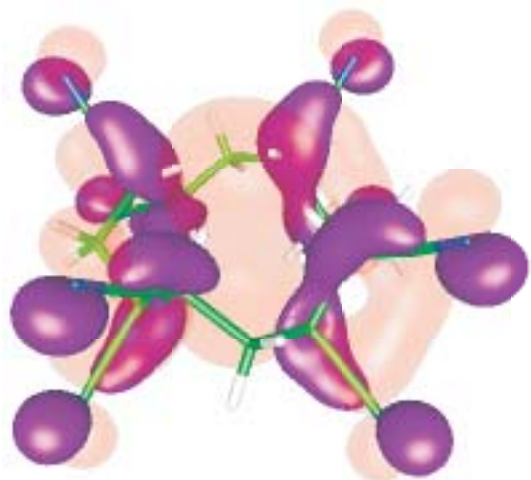

(b)

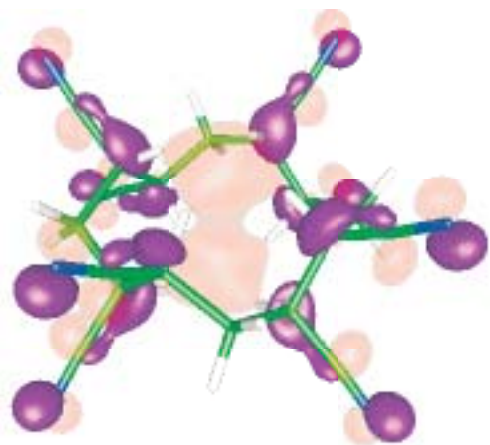

(c)

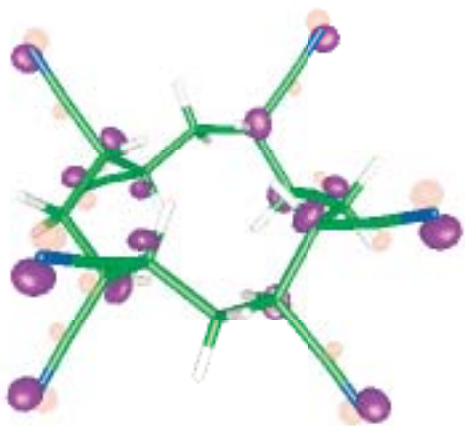

**Figure 9S; Shkrob & Sauer**

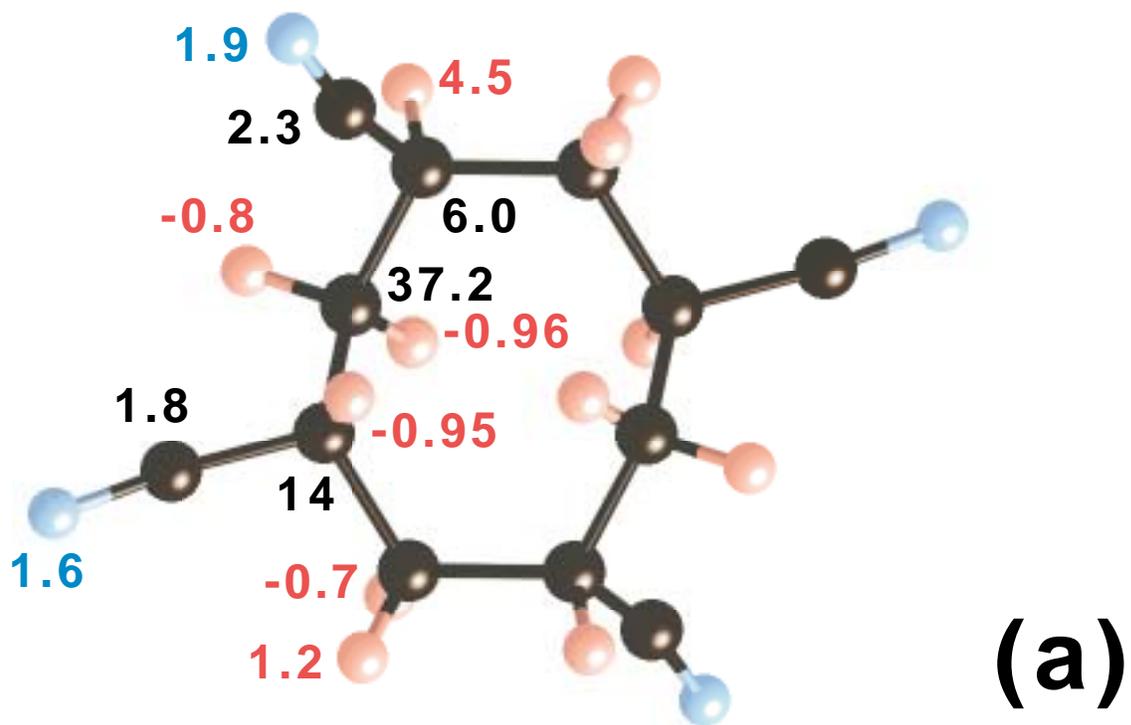

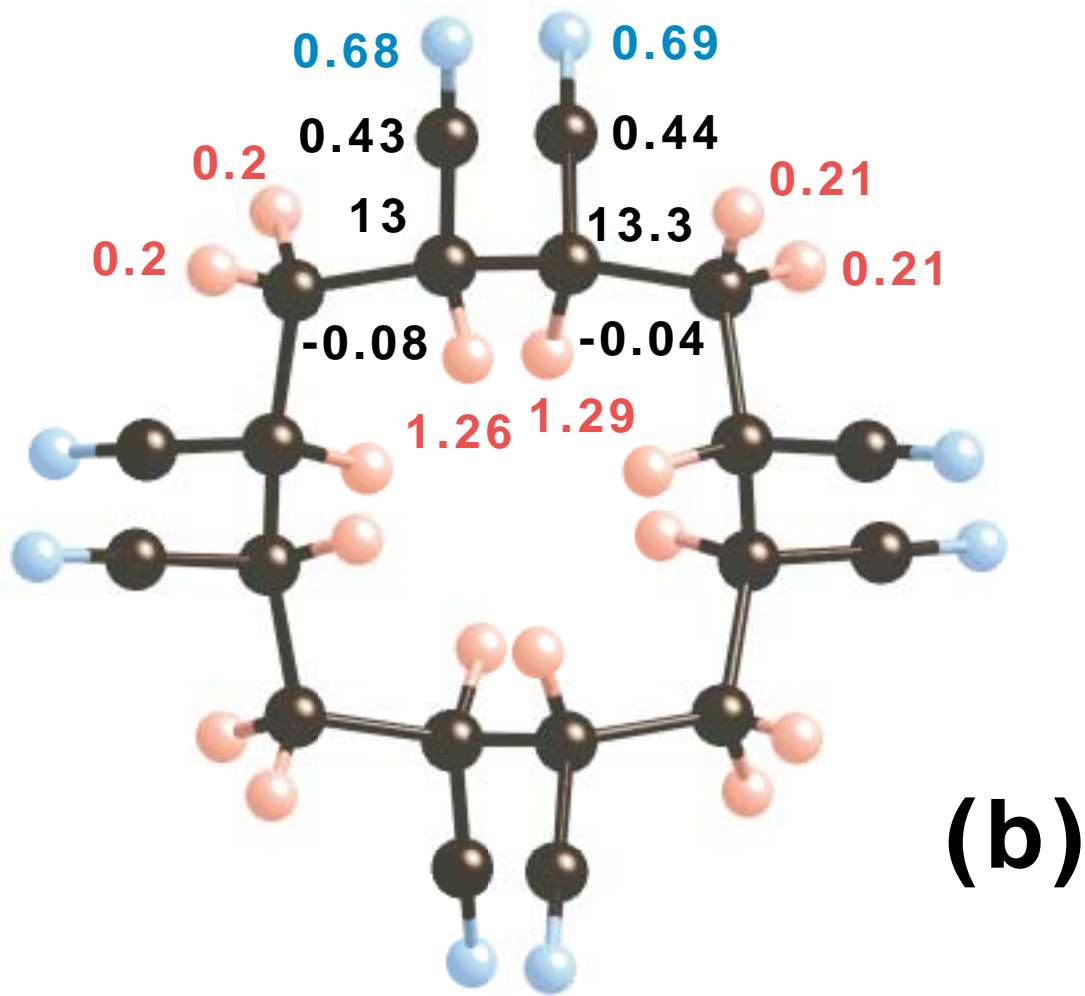

Figure 10S; Shkrob & Sauer